\documentstyle[sprocl,twoside,psfig]{article}

\def\citebk#1{\hspace{0.9mm}\raisebox{-1.85mm}[0mm][0mm]
  {\Large\cite{#1}}\hspace{-0.1mm}}

\def\Journal#1#2#3#4{{#1} {\bf #2}, #3 (#4)}

\def\NCA{\em Nuovo Cimento}

\def\NPB{{\em Nucl. Phys.} B}
\def\PLB{{\em Phys. Lett.}  B}
\def\MPLA{{\em Mod. Phys. Lett.}  A}
\def\IJMP{{\em Int. J. Mod. Phys.}  A}
\def\PRL{\em Phys. Rev. Lett.}
\def\JHEP{\em J. High Energy Phys.}
\def\PRD{{\em Phys. Rev.} D}
\def\PRB{{\em Phys. Rev.} B}

\def\be{\begin{equation}}
\def\ee{\end{equation}}
\def\bea{\begin{eqnarray}}
\def\eea{\end{eqnarray}}
\def\beq{\begin{eqnarray}}
\def\eeq{\end{eqnarray}}
\def\eval#1{\left\langle#1\right\rangle}
\def\bA{{\bf A}}
\def\bm{{\bf m}}
\def\bn{{\bf n}}
\def\br{{\bf r}}
\def\bs{{\bf s}}
\def\bk{{\bf k}}
\def\bx{{\bf x}}
\def\by{{\bf y}}
\def\bz{{\bf z}}
\def\d{\partial}
\def\dd{\partial}
\def\ep{\epsilon}

\begin{document}


\title{MONOPOLES, VORTICES AND STRINGS: CONFINEMENT AND
 DECONFINEMENT IN 2+1 DIMENSIONS AT WEAK COUPLING}

\author{IAN I. KOGAN }  

\address{Department of Physics, Theoretical Physics, 1 Keble 
Road,\\
Oxford OX1 3NP, UK}
\author{ALEX KOVNER }
\address{ Department of Mathematics and Statistics,
 University of Plymouth, 2 Kirkby Place, Plymouth PL4 8AA, 
UK}

\maketitle

\abstracts{We consider, from several complementary 
perspectives,
the physics of confinement and deconfinement in the 2+1
dimensional Georgi-Glashow model. Polyakov's monopole plasma 
and
't Hooft's vortex condensation are discussed first. We then 
discuss
the physics of confining strings at zero temperature. We 
review
the Hamiltonian variational approach and show how the linear
confining potential arises in this framework. The second 
part of
this review is devoted to study of the deconfining phase
transition. We show that the mechanism of the transition is 
the
restoration of 't Hooft's magnetic symmetry in the deconfined
phase. The heavy charged $W$ bosons play a crucial role in 
the
dynamics of the transition, and we discuss the interplay 
between
the charged $W$ plasma and the binding of monopoles at high
temperature. Finally we discuss the phase transition from 
the
point of view of confining strings. We show that from this 
point
of view the transition is not driven by the Hagedorn 
mechanism
(proliferation of arbitrarily long strings), but rather by 
the
``disintegration" of the string due to the proliferation of 
$0$
branes.  }

  \vspace{1cm}

\tableofcontents
\newpage

\section{Introduction}

Study of the confining regime in QCD is one of the most 
elusive
and complicated subjects in modern particle physics. The 
theory is
strongly interacting, and our arsenal as of today does not 
contain
any reliable analytical tools to tackle strongly interacting
quantum field theory. The situation is exacerbated in QCD by 
the
fact that confinement is not the only nonperturbative 
phenomenon
that plays out in the QCD vacuum. The spontaneous breaking 
of
approximate chiral symmetry is the other important player.
Although we believe that in principle the two phenomena are 
quite
distinct, in the framework of QCD they are not easily 
separable.
It is thus very important to have a simpler setting where 
one can
understand the physics of confinement without having to 
worry
about other complicated phenomena.

Thankfully, we do know a model, or rather a class of models, 
which
exhibit confinement but are free from other complications of 
QCD.
Those are weakly interacting confining gauge theories in 2+1
dimensions. They were first studied by 
Polyakov\,\cite{polyakov}
almost twenty years ago. By now we understand the confining
physics of these theories very well and have several 
perspectives
on it.  These  results  have been confirmed by numerical 
methods
of lattice gauge theory.\cite{lattice} Moreover in the last 
couple
of years the finite temperature behavior of these theories 
has
also been well understood, including the details of the
deconfining phase transition.

Our aim in this review is to discuss in detail, and from 
several
complementary points of view, the physics of these theories. 
We
start in Sec. 2 with the description of the simplest of 
these
models  ---  the SU$(2)$ gauge theory with the adjoint Higgs 
field  --- 
the Georgi-Glashow model. We describe Polyakov's derivation 
of
confinement in this model based on the famous monopole 
plasma
mechanism. In Sec. 3 we explain how confinement arises from 
the
point of view of condensation of magnetic vortices. We show 
how
the two pictures are closely related, but also point to some
subtle differences between them and their complementary 
aspects.
This has been discussed in detail\,\cite{kovnerr} in a 
previous 
volume in this
series, and so our discussion will be brief.
Section 4 is devoted to the discussion of Polyakov's 
confining
string. We explain how the rigidity of the mathematical 
string
arises from the absence of ``stringy" degrees of freedom in 
the
ultraviolet. We also explain why the $W$ bosons play the 
role of
D0 branes in the theory of the confining string. In Sec. 5 
we
show how the physics of confinement can be understood using
Hamiltonian methods. We present a gauge invariant 
Hamiltonian
variational calculation of the ground state wave function as 
well
as of the potential between external sources. Section 6 is 
devoted
to a detailed discussion of the deconfinement phase 
transition and
the calculation of critical exponents. In Sec. 7 we discuss 
the
properties of the vortex correlation functions and of the 
monopole
interactions in the high temperature phase. We show that the
monopoles are linearly ``confined", while the vortex 
correlators
decrease exponentially with distance. Section 8 discusses 
the same
phase transition from the perspective of the confining 
string. It
is shown that the would be ``monopole binding" transition
corresponds to the Hagedorn transition in the string 
language. The
actual mechanism of the deconfining transition is the
disintegration of the string due to proliferation of D0 
branes.
Section 9 presents the generalization of various aspects of 
the
preceding discussion to the SU$(N)$ Higgs model. Finally, 
Sec.~10 
contains some concluding remarks.

\section{The monopole plasma}
We start our discussion of confinement in 2+1 dimensions by
recapitulating the original derivation of Polyakov. Consider 
the
2+1 dimensional SU$(2)$ gauge theory with a scalar field in 
the
adjoint representation,
 \be S= -{1\over 2}\int d^3x
\mbox{tr}\left(F_{\mu \nu}F^{\mu \nu}\right) + \int d^3x
\left[{1\over 2}(D_\mu h^a)^2 +{\lambda \over  4}(h^a h^a - 
v^2)^2
\right]. \label{model1} 
\ee 
We adopt the notation   $$A_\mu =
{i\over 2} A^a_\mu \tau^a\,,\quad F_{\mu\nu} = \dd_\mu A_\nu  
-\dd_\nu
A_\mu +g[A_\mu, A_\nu] \,,\quad h = {i\over 2} h^a 
\tau^a\,,$$
and
$$ D_\mu h
= \dd_\mu h + g[A_\mu, h]\,.$$ 
The trace is normalized 
as
$$\mbox{tr}(\tau^a \tau^b)= 2 \delta^{ab}\,.$$

We concentrate on the weakly coupled regime $v\gg g^2$. 
Since the
coupling is weak one can try one's hand first at 
perturbation
theory. Perturbatively the SU$(2)$ gauge group is broken to 
U$(1)$
by the large expectation value of the Higgs field. The 
photon
associated with the unbroken subgroup is massless whereas 
the
Higgs and the other two gauge bosons $W^\pm$ are heavy with 
the
masses
\begin{equation}
M^2_H= 2\lambda v^2, \hskip 1.5 cm M^2_W=g^2v^2.
\end{equation}
Thus, perturbatively the theory behaves very much like
electrodynamics with spin one charged matter. However, this 
is not
the whole story. It is known rigorously\,\cite{frsh}  that 
the
weakly coupled regime in this model is not separated by a 
phase
transition from the confining regime, which one naturally
identifies with the vanishing Higgs expectation value. In 
the
confining regime the spectrum of the theory is certainly 
massive,
and the charged particles are linearly confined. Thus, there 
must
be some nonperturbative effects that bring about these two 
changes
even at weak coupling. The nonperturbative effects in 
question are
the monopoles, as first discussed by 
Polyakov.\cite{polyakov}

It is easy to see that the Euclidean equations of motion of 
the
model (\ref{model1}) have monopole like solutions,
 \beq
D_\mu
F^{\mu\nu}  = g[h,D^\nu h], ~~~~~~
  D_\mu D^\mu h =
-\lambda(h^ah^a - v^2) h \,.
\label{EulerEq1} 
\eeq
 The
perturbative vacuum manifold is SO$(3)/{\rm U}(1)$ and this 
has a
non-zero second homotopy group, 
$\Pi_2({\rm SO}(3)/{\rm U}(1))= Z$. Thus, the
equations admit topologically nontrivial solutions with 
winding of
the Higgs field at the space-time infinity.
 The ``elementary"
configuration of this type is the well-known 
`t Hooft-Polyakov monopole,
\bea
h^a(\vec{x}) & = &\hat x^a h(r)\,, \nonumber \\
gA^a_\mu(\vec{x}) & = &{1\over r} \epsilon_{a\mu
\nu}\hat{x}^\nu(1-\phi(r)\,,
 \label{configuration1}
\eea
where $\hat x^a = x^a/r$. The non-Abelian field strength is
\bea
gF_{\mu\nu}^a =
 {1\over r^2}
  \ep_{\mu\nu b}  \hat x^a \hat x^b (\phi^2  -1)
+ {1\over r} (\ep_{a\mu\nu} -  \ep_{\mu\nu b}  \hat x^a \hat 
x^b )
\phi',
 \label{non-abelian field strength}
\eea where the prime denotes differentiation with respect to 
the
3D radial coordinate $r$. The boundary conditions satisfied 
by the
various fields are $h(0)= 0$ , $h(\infty)= v$, $\phi(0)= 1$ 
and
$\phi(\infty)= 0$. The magnetic field far away from the 
origin is
radial in the color space, and has a typical monopole-like
behavior, dying away as $1/r^2$. In fact, the Abelian field
strength

\be F_{\mu\nu} = \hat h^a F_{\mu\nu}^a - {1\over g} 
\epsilon_{abc}
             \hat h^a(D_\mu \hat h)^b (D_\nu \hat 
h)^c\label{abelian}
\ee 
where $$\hat h^a={h^a\over |h|}\,,$$ is precisely that of a
magnetic monopole everywhere except a small region around 
the
origin  ---  the monopole core. The core size is of the 
order of the
vector meson mass. At weak coupling the action of the 
monopole
solution is very large \be S_{\rm monopole}={2\pi M_W \over 
g^2}
\epsilon\left({M_H\over M_W}\right) \,,
\ee 
where $$\epsilon\left({M_H\over 
M_W}\right)$$ is a
slowly varying function such that $ 1\leq \epsilon \leq 
1.787$
(further details can be found in Ref.~\citebk{Prasad}), and $\epsilon(\infty)=1$. Due to 
this 
large value
of the action the contribution of the monopoles to the 
partition
function is very small. Thus, typical configurations contain 
only
very low density of monopoles and this dilute ``monopole 
plasma"
can be treated reliably in the fugacity expansion.

The monopoles interact via the three-dimensional Coulomb
potential. The classical action for a configuration that 
contains
far separated monopoles and anti-monopoles is 
\be S =
\frac{2\pi}{g^2} \sum_{a \neq b} \frac{q_a q_b}{|\vec{r}_a -
\vec{r}_b|}
 + {2\pi M_W \over g^2}
\epsilon\left({M_H\over M_W}\right)\sum_{a} q^{2}_{a} \ee where $q_{a} 
= \pm
1$. The monopole fugacity is determined as 
\be \xi=\exp\left[-
{2\pi
M_W \over g^2} \epsilon\left({M_H\over M_W}\right)\right]\,. 
\label{fugacity}
\ee

 Even though this fugacity is tiny, the monopoles have a 
decisive
 effect on the low-energy properties of the theory.
 Since the $W$ bosons are very heavy, one can neglect their
 effects with good accuracy. In this approximation, the 
partition
 function of the theory factorizes into the
 partition function of the massless photon and that of the 
monopole
plasma. The two factors can be combined into the partition
function of a ``dual photon" field.
 Following Polyakov's 
Derivation\,\cite{polyakov} we start from the  statistical 
sum  
$Z_M$ for
Coulomb plasma with $N_{+}$ positive and $N_{-}$ negative 
charges
with equal fugacities $\xi$, which is given by \be 
Z_{\rm plasma} =
\sum_{N_{+}, N_{-}}\int \prod_{a = 1}^{N_{+}} d^3 
\vec{r}_{a}
  \prod_{b=1}^{N_{-}} d^3 \vec{r}_{b}\mu^{3(N_{+}+N_{-})}
\frac{1} {N_{+}!~N_{-}!} \exp\left(-S\{a,b\}\right) \ee 
where
$\vec{r}_{a}$ and  $\vec{r}_{b}$ are coordinates for 
positive and
 negative charges respectively.
The dimensionful constant $\mu$ is necessary to convert the
summation over the monopole coordinates into the 
integration. It
is nonuniversal, but clearly is related to the shortest 
distance
scale at which the present approximation is valid, namely 
the size
of the monopole core.
 In Polyakov's calculation $\mu$ is determined to be
 \be \mu^3\propto{M_W^{7/2}\over g}\,.\ee

Using the fact that the Coulomb potential is the inverse of 
the
Laplacian operator, this can be written as\,\cite{Coleman}
\begin{eqnarray} 
&&Z_{\rm plasma} =Z^{-1}_0 \int 
D\phi\sum_{N_{+},
N_{-}}\int \prod_{a = 1}^{N_{+}} d^3 \vec{r}_{a}
  \prod_{b=1}^{N_{-}} d^3 \vec{r}_{b}
\frac{(\mu^3\xi)^{(N_{+}+N_{-})}} {N_{+}!~N_{-}!}\\
&& \exp\left(-\int dx{g^2\over 32\pi^2} 
(\partial_\mu\phi)^2-
\sum_{a,b}i\phi(x)[(\delta(x-x_a)-\delta(x-
x_b))]\right).\nonumber\label{plas}
\end{eqnarray} Here $Z_0$ is the partition function of a 
single
massless field. In the dilute gas approximation, that is 
allowing
at most one monopole or antimonopole at each spatial point, 
the
summation over the coordinates is easily performed and one 
obtains
for the partition function of the theory

\be Z = \int D\phi  \exp\left[-\int d^3 x {\cal 
L}_{\rm eff}\right] \ee 
with
the effective Lagrangian 
\be {\cal L}_{\rm eff} = {g^2\over32
\pi^2}(\d_{\mu}\phi)^2  - \xi\mu^3\cos \phi \,.
\label{sine-Gordon}
\ee 
Note that the factor $Z_0^{-1}$ in Eq.~(\ref{plas}) 
cancelled
against the contribution of a free massless photon of the
Georgi-Glashow model.

One striking effect of the monopole plasma is explicit in 
this
representation. The field $\phi$ which in  
Eq.~(\ref{sine-Gordon})
represents the photon, is not massless. Rather it has an
exponentially small mass, determined by the monopole 
fugacity,
\be M_\gamma^2= {16 \pi^2 \mu^3\xi\over{g^2}} \,.
\label{Mgamma}
\ee
The other effect is confinement, or area law for the 
fundamental Wilson loop.

Let us  calculate the expectation value of the Abelian 
Wilson loop
\be W(C) =\left\langle\exp\left(ig\int_{\Sigma} d^2 x 
B(x)\right)\right\rangle\,,
\ee 
where
$B$ is a component of the Abelian magnetic field 
(\ref{abelian}) and $\Sigma$ is a surface bounded by the 
curve
$C$.

To calculate the expectation value, we insert this 
exponential
factor into the path integral for the partition function of 
the
model and use the fact that for a monopole plasma 
configuration
\be \exp\left(ig\int_{\Sigma} d^2 x B(x)\right) = 
\exp\left(i\int
d^3 x \eta(x) \rho(x)\right). \ee Here \be
\eta(x)=\int_{\Sigma}{{(x-y)_{\mu} \over | x-y|
^{3}}d^{2}\sigma_{\mu}(y)} \ee is the flux (normalized at 
$4\pi$)
through the surface $\Sigma$  from  a  monopole located at
 the point $x$,  and $$ \rho(x) = \sum_{a,b} [\delta(x-x_a)-
\delta(x-
x_b)]$$
 is the density of the magnetic charge in the plasma.
As far as the summation over the monopole coordinates is 
concerned
this is exactly the same type of term as the linear term in 
$\phi$
in the exponential in Eq.~(\ref{plas}). Thus, the summation 
can
again be performed, and one obtains \be W(C) = \frac{1}{Z}
 \int D\phi  \exp\left[-\int d^3 x {\cal 
L}_{\rm eff}(\eta)\right]
\ee where \be {\cal L}_{\rm eff}(\eta) = {g^2\over32 
\pi^2}(\d_{\mu}\phi)^2
- {M_\gamma^2 g^2\over 16\pi^2}\cos(\phi + \eta) \,.
\ee

For large contours $C$ the path integral over $\phi$ can be
calculated now in the steepest descent approximation. It is
convenient to change variables to $\Phi=\phi+\eta$. The 
integral
then is determined by the solution of classical equations of
motion
 \be
\partial^2(\Phi-\eta) = M_\gamma^2 \sin(\Phi)\,.
\ee 
The nature of this solution is easy to understand. Let 
us for
definiteness imagine that the surface $\Sigma$ lies in the 
1-2 plane. The Laplacian of $\eta$ vanishes everywhere in 
space, except on the surface $\Sigma$, where it is equal to 
the transverse derivative of the delta function with the 
coefficient
$2\pi$,
 \be\partial^2\eta(x)=2\pi\partial_z\delta^3(x-
\Sigma). 
\ee
 Thus, the
field $\Phi$ 
satisfies the free sine-Gordon equation everywhere in space, 
except on the surface 
$\Sigma$. The presence of 
$\eta$
imposes the boundary condition that this particular solution 
must
be discontinuous across $\Sigma$ with a discontinuity of 
$2\pi$.
Such a solution is easily constructed from a domain wall 
solution
of the sine-Gordon theory. At negative $z$ the solution 
$\Phi$ is
just the domain wall, so that at $z=0+$ it reaches the value
$\pi$. It then jumps discontinuously to $-\pi$ on the other 
side
of the surface, $\Phi(z=0-)=-\pi$ and rises back to zero at 
large
positive $z$ again following the domain wall solution. The 
action
of such a solution is obviously just equal to the action of 
a
domain wall in the sine-Gordon model. It is proportional to 
the
total area of the surface $\Sigma$, and the coefficient
parametrically is given by $g^2M_\gamma$. Thus, we find

 \be W(C)  \sim
\exp\left( - \sigma S  \right) \ee where $S$ is the minimal 
area
and $\sigma \sim g^2 M_\gamma$. The Wilson loop therefore 
has an
area law and the theory is confining.

Thus, the monopole plasma accounts nicely for the finite 
mass of
the photon as well as for the area law of the Wilson loop.

\section{The vortex condensation}

A complementary view of confinement in this theory is the
spontaneous breaking of the magnetic $Z_2$ symmetry due to 
the
condensation of magnetic vortices.\cite{'t Hooft,samuel}
This approach has been reviewed in Ref. \citebk{kovnerr} and 
therefore we
will be brief here.

The Georgi-Glashow model has a global $Z_2$ symmetry. This
symmetry is not recognizable as a Noether type symmetry in 
the
Lagrangian (\ref{model1}), but rather is akin to 
``topological"
symmetry. It is generated by the total magnetic flux. 
Ignoring for
the moment monopole effects,  the Abelian magnetic field
(\ref{abelian}) is a conserved current
\begin{equation}
\tilde F^\mu={1\over 
4}\epsilon_{\mu\nu\lambda}F_{\nu\lambda}.
\end{equation}
However, due to the monopole contributions (see Ref. 
\citebk{Kovner}),
this symmetry is anomalous, and only the $Z_2$ subgroup is
conserved. The non-anomalous $Z_2$ magnetic symmetry group 
is
generated\,\cite{Kovner1} by the large spatial Wilson loop 
which
encloses the system,
 \be
 W(C \rightarrow \infty ) = 
\mbox{exp}\left(
{ig\over 2}\int d^2 x B(x)\right)\,. 
\ee 
The order parameter for this 
$Z_2$
transformation is the operator that creates\,\cite{'t 
Hooft,Kovner} 
an elementary magnetic
vortex of flux $2\pi/g$, \be V(x) =
\mbox{exp}{{2\pi i\over g}} \int_C \epsilon_{ij}\hat h^a 
E^a_j(x).
\label{v}\ee This operator is local, gauge invariant and a 
Lorentz
scalar. Together with the spatial Wilson loop it forms the
order-disorder algebra,
 \be W(C \rightarrow \infty )V(x) =
 - V(x) W(C \rightarrow\infty)\,.
\ee
The vortex operator condenses in the vacuum and its 
expectation
value is determined\,\cite{kovnerr} by the gauge coupling 
constant
$\langle V\rangle^2=g^2/8\pi^2$. 

Since the anomalous effects that break the U$(1)$ magnetic
symmetry down to $Z_2$ at weak coupling are small, one can
construct the low-energy effective theory Lagrangian in 
terms of
the relevant order parameter. It is given by the $Z_2$ 
invariant
theory of the vortex field $V$,
 \be {\cal{L}}_{\rm eff}= 
\dd_\mu V\,
\dd^\mu V^* - \lambda \left( V\, V^* - {g^2\over 
8\pi^2}\right)^2 + 
{M^2\over
4} \{ V^2 + (V^*)^2\}. 
\label{lowlagrangian} 
\ee
 The vortex self-coupling in this effective Lagrangian can 
be  determined as\,\cite{Kovner2}
\be
\lambda={2\pi^2M_W^2\over g^2}.
\label{couplings2}
\ee

Note that as a low-energy Lagrangian, Eq.
(\ref{lowlagrangian}) is
indeed consistent with Eq.~(\ref{sine-Gordon}). At weak 
gauge
coupling, the quartic coupling $\lambda$ is very large. In 
this
nonlinear $\sigma$-model limit the radius of the field $V$ 
is
therefore frozen to its expectation value. The only relevant
degree of freedom is the phase \be V(x) = {g\over 
{\sqrt{8}\pi}}
e^{i\chi}. \ee Substituting this into 
Eq.~(\ref{lowlagrangian}) one
indeed obtains precisely Eq.~(\ref{sine-Gordon}) with the
identification $\phi=2\chi$. Thus, Polyakov's dual photon is
nothing but the pseudo-Goldstone boson of the Lagrangian 
(\ref{lowlagrangian}).

This relation can be inferred directly from the definition 
of $V$ in
Eq.~(\ref{v}) and the monopole plasma partition function
(\ref{plas}). Equation (\ref{plas}) relates the field 
$\phi$ 
with the
monopole density as 
\be {g^2\over 16\pi^2}\phi(x)=i\rho ={g\over
4\pi}\partial_\mu\tilde F^\mu\,.
\ee 
The factor $i$ has disappeared in
the second equality since the field strength is now  in 
Minkowski
rather than Euclidean space. The factor ${g\over 4\pi}$ is 
due to
the fact that the minimal magnetic monopole charge is twice 
the
fundamental value allowed by the Dirac quantization 
condition.
This immediately gives 
\be 
\phi(x)={4\pi\over 
g}{\partial_\mu\over
\partial^2}\tilde F^\mu={4\pi\over g}\int_{x_0}^x dz_\mu 
\tilde
F^\mu(z).
\ee 
The integral on the right-hand side does not 
depend
(to leading order in fugacity) on the choice of the 
integration
curve. Choosing it to be at equal time and comparing to
Eq.~(\ref{v}), we get 
\be\phi=2\chi\,.\ee 
Thus, the non-vanishing
photon mass is a natural consequence of the fact that the 
magnetic
symmetry is discrete and not continuous, and thus the photon 
is a pseudo-Goldstone boson.

The calculation of the spatial Wilson loop in this framework
closely follows the calculation in the previous subsection. 
For
details we refer the reader to Ref. \citebk{kovnerr}.

Thus, as far as the description of the low-energy phenomena 
is
concerned, the vortex approach is equivalent to the monopole
plasma approach. The one conspicuous difference between them 
is
that in the vortex low-energy approach the field $\chi$ is 
treated
as a phase, while in the monopole plasma context $\phi$ is 
an
ordinary real field which takes values on a straight line. 
As a
consequence the Lagrangian (\ref{lowlagrangian}) allows
configurations with quantized discontinuities of the field 
$\chi$.
In particular, the vortex configurations where the field 
$\chi$
winds around some point are allowed in the description of
Eq.~(\ref{lowlagrangian}). In such a vortex configuration 
the field
$\chi$ is discontinuous along an infinite cut. In the 
framework of
the effective theory, Eq.~(\ref{lowlagrangian}), such a 
quantized
discontinuity does not contribute to the energy. Polyakov's
effective Lagrangian (\ref{sine-Gordon}), on the other 
hand,
does not allow such vortex configurations. Since the field 
$\phi$
in it is not treated as a phase, the cut does contribute to 
the
kinetic energy term in Eq.~(\ref{sine-Gordon}), and this
contribution is both ultraviolet and infrared divergent. The
ultraviolet divergence is of no importance by itself, since 
the
theory is defined with the intrinsic ultraviolet cutoff 
$M_W$. 
However, the
infrared divergence indisputably puts the description of 
such
vortex-like configurations beyond Eq.~(\ref{sine-Gordon}).

As it turns out this difference can sometimes be quite
significant. The point is that these vortex-like 
configurations of
the field $V$ are the low-energy incarnation of the charged 
states
of the theory, the $W$ bosons.

As explained in Ref. \citebk{kovnerr}, the electric current 
is 
identified
with the topological current in Eq.~(\ref{lowlagrangian}),
\begin{equation}
{\frac{g}{\pi }}J_{\mu }=i\epsilon _{\mu \nu \lambda 
}\partial
_{\nu }(V^{\ast }\partial _{\lambda }V).
\end{equation}
Thus, charged states carry unit winding number of $\chi$. 
The
$W^+$ ($W^-$) boson is a state with positive (negative) unit
winding of $\chi$.

Thus, the physics of the charged states cannot be discussed 
within
Eq.~(\ref{sine-Gordon}), but can be considered within
Eq.~(\ref{lowlagrangian}). Of course since the charged 
states are
confined in this theory, the energy of an isolated charged 
state
is linearly infrared divergent. The mechanism of this 
divergence,
as well as the formation of the electric flux tube, is 
discussed
in detail in Ref. \citebk{kovnerr}. As we will see in later 
sections,
proper treatment of the charged states is crucial for the
understanding of the dynamics of the deconfinement phase
transition and thus it is worthwhile to be careful on this 
point.
We note that this is not the only instance in which the 
charged
states play an important role. For example, as discussed in 
Ref.
\citebk{greensite}, the presence of these states leads to 
the
breaking of the string of charge two, which in Polyakov's
effective Lagrangian (\ref{sine-Gordon}) is strictly 
stable.

In fact, to properly account for the large mass of the 
charged
particles, the Lagrangian (\ref{lowlagrangian}) must 
be
slightly modified, since as it stands it underestimates 
their
energy. The core energy of the soliton of $V$ should be 
equal to
the mass of the charged vector boson $M_W$, since this is 
the
lightest charged excitation in the theory. On the other 
hand, the
core mass of a soliton in the Lagrangian 
(\ref{lowlagrangian}) 
is given by the ultraviolet contribution of the
Coulomb potential. With the cutoff of order $M_W$, this is 
of
order $g^2 \ln M_W/g^2$. This is not surprising, since the 
mass of
the $W$ boson indeed comes from the distances of the order 
of its
Compton wavelength, and is thus well inside the ultraviolet 
cutoff of 
the
effective theory. The situation can be improved by adding to 
the
effective Lagrangian a higher derivative term of the Skyrme 
type
\be \delta L=\Lambda(\epsilon_{\mu\nu\lambda}\partial_\nu 
V^*
\partial_\lambda V)^2,
\label{skyrme} \ee with \be
\Lambda\propto{\frac{1}{g^4}}{\frac{1}{M_W}}. \ee The extra 
term
Eq.~(\ref{skyrme}) does not affect the photon mass nor the 
value of
the string tension. The density of this action does not 
vanish
only at the points where the phase of $V$ is singular  ---  
that is
the points of winding. For a closed curve $C$ of length $L$ 
which
carries the winding, the contribution of this extra term to 
the
action is $M_WL$, which is precisely the action of the 
world line
of a massive particle of mass $M_W$.

Thus, Polyakov's monopole plasma picture and 't Hooft's 
magnetic
vortex condensate picture fit very neatly together and 
provide a
fairly exhaustive understanding of confinement in the
Georgi-Glashow model. Of course, both monopoles and vortices 
are
not physical excitations of the Georgi-Glashow theory. What 
one
would like to see is the confining string, the tube of 
electric
flux. It is believed that in 3+1 dimensional QCD at large 
$N_c$
these strings interact weakly and perhaps are described by a 
free
string theory. Although one does not expect the same to be 
true in
2+1 dimensions at $N_c=2$, the handle we have on the present 
model
gives us a unique possibility to try and derive the relevant
string theory. In the next section, therefore, we explain 
how one
can derive in the long wavelength approximation the theory 
of the
confining string. This theory turns out to have some quite
peculiar features, which we will discuss.

\section{Confining strings}
In 1996 Polyakov   proposed\,\cite{POL96}  the derivation of 
the 
so-called
confining string action explicitly from compact QED, which 
is 
equivalent to the Georgi-Glashow model in
the low-energy approximation of Eq.~(\ref{sine-Gordon}). 
This
action was studied further in 
Ref.~\citebk{confiningstringaction} 
and it
was found to describe a rigid 
string.\cite{Rigid1,confinigstringreferences}

We will first discuss the original  derivation due to 
Polyakov
which uses the  Kalb-Ramond field.\cite{KALB} We will also 
give a
more intuitive and simple
 derivation which  uses the direct
correspondence between field configurations and the 
summation over
closed string world sheets. An advantage  of this derivation 
is
that it makes explicit an important and quite unusual 
property of
the confining string, namely that the fluctuations of this 
string
with large momenta (larger than the inverse thickness of the
string) do not cost energy. This is a dynamical extension of 
the
so called zigzag symmetry introduced by 
Polyakov.\cite{POL96} We
also show that the heavy charged particles of the 
Georgi-Glashow
model ($W^\pm$ bosons) appear as a certain kind of $0$ 
branes in
the string description.

Our starting point will be an expression for the Wilson loop
discussed
 earlier,
\be W(C) = \frac{1}{Z}
 \int D\phi  \exp\left[-\int d^3 x {\cal 
L}_{\rm eff}(\eta)\right],
\ee where \be {\cal L}_{\rm eff}(\eta) = {g^2\over32 
\pi^2}(\d_{\mu}\phi)^2
- {M^2 g^2\over 16\pi^2}\cos(\phi + \eta). 
\label{wilsonl}\ee

Note that the effective theory at $\eta=0$ has more than one
vacuum state. In particular, $\phi=2\pi n$ for any integer 
$n$ is
the classical ground state. Therefore, the classical 
equations of
motion have wall-like solutions  ---  where the two regions 
of space,
say with $\phi=0$ and $\phi=2\pi$ are separated by a domain 
wall.
The action density per unit area of such a domain wall is
parametrically
\begin{equation}
\sigma\propto g^2M.
\end{equation}
As we discussed in Sec. 2, the fundamental Wilson loop, when
inserted into the path integral, induces such a domain wall
solution, with the result
\begin{equation}
\langle W(C)\rangle =\exp\{-\sigma S\},
\end{equation}
where $S$ is the minimal area subtended by the contour $C$. 
Thus,
the string tension is precisely equal to the wall tension of 
the
domain wall. This domain wall is identified with the 
world sheet of
the confining QCD string.

\subsection{The Kalb-Ramond way}

 In order to find a  string representation for $W(C)$  one  
can 
rewrite
$W(C)$ in a different form, introducing a new auxiliary
Kalb-Ramond field $B_{\mu\nu}$, \be W(C)  =  \int 
DB_{\mu\nu}D\phi
\exp[-S(\phi,B)] \ee with the  action \bea S(\phi,B) & = 
&\int d^3
x
 \left[{ 16\pi^2 \over g^{2}}B_{\mu \nu}^{2}+i 
\epsilon_{\mu\nu\rho}
\partial_{\mu}\phi B_{\nu\rho}
+ {M^2 g^2\over 16\pi^2} (1-\cos\phi)\right] \nonumber \\
 & + & i\int_{\Sigma_{C}}{B_{\mu\nu}d\sigma_{\mu\nu}}.
\eea
Integration over $B$ brings one back to the initial 
expression for
 Wilson loop Eq.~(\ref{wilsonl}).
  We may choose, however, to integrate first over  $\phi$. 
In that 
case
   we obtain an  effective
action for the massive $B$ field,
 \be W(C) =\int{DB 
\exp\left[\int
dx \left( { 16\pi^2 \over g^{2}}B_{\mu \nu}^{2}
+f(H)\right)\right] \exp{i\int_{\Sigma_{C}}{Bd\sigma}}} \,,
\ee 
where
$H=dB$  is the field strength and the function $H$ is given 
by
(for details see Ref. \citebk{POL96}) 
\be f(H) = \left[ H\arcsin{\left({H 
\over
M^{2}}\right)}\right] -\sqrt{M^{4}-H^{2}} \,.
\ee
 The multivaluedness of this action reflects the periodicity 
in
 $\phi$ mentioned earlier.
 Expanding in powers of $B$ and  limiting ourselves to 
quadratic 
terms,
 the  effective  action takes the form, \be S(B)= { 16\pi^2
\over g^{2}}\int d^3 x \left[ B_{\mu \nu}^{2} + 
\frac{const}{M^2}
H_{\mu\nu\rho}^2 \right]. \ee For large loops one gets \be 
W(C)
\sim \exp\left(-F(C,S))\right) \ee
 with the following local expansion for the string action  
$F$:
\be F = c_1 g^2 M \int{d^{2}\xi \sqrt{g}}
+c_{2}g^{2}M^{-1}\int{d^{2}\xi ( \nabla t_{\mu \nu} )^{2}
\sqrt{g}}  + \ldots \ee
 where
\be g_{ab} = \partial_a \vec{x} \partial_b \vec{x}, ~~~~~
t_{\mu\nu}=\epsilon ^{ab}\partial_{a} x_{\mu} \partial_{b}
x_{\nu}. \ee
 This is the action of the rigid string which also can be 
written as
(see Ref. \citebk{confinigstringreferences} and references 
therein)
\begin{equation}
 F = \int d^2 \xi \sqrt{g} g^{ab} D_a \vec{x} \left( \sigma 
- s 
D^{2} +
..\right) D_b \vec{x}\label{stringa}
\end{equation}
where $g$ and $D_a$ respectively are the determinant and the
covariant derivative with respect to the induced  metric $ 
g_{ab}
=
\partial_a \vec{x}
\partial_b \vec{x}$ on the embedded world  sheet 
$\vec{x}(\xi_1,
\xi_2)$. Note that the rigidity is controlled by the second 
term
and  one can show that the stiffness $s$  is parametrically  
of
order
\begin{equation}
 s \propto{\sigma\over M^2}.
\end{equation}

The stiffness $s$ in this expression is actually negative, 
but the
system is stable due to higher order terms. Note that since
$\sigma\propto g^2M\gg M^2$, the rigidity is very large, and 
is
effective on the distance scales much larger {\it 
parametrically}
than the ``natural'' string scale given by the string 
tension.

The action (\ref{stringa}) explicitly depends on the 
surface
$\Sigma$ , while originally it was introduced as an 
unphysical
object. The  reason for this  is the multivaluedness of the  
full
action $S(B)$. As was shown in Ref.~\citebk{POL96} the 
surface independence of $\langle W(C)\rangle$  is
 restored  after summation   over all
possible  surfaces.

\subsection{The confining string for pedestrians}
 Let us now derive the action of a fluctuating string in a 
More direct way.\cite{martin}
We work with the Polyakov effective Lagrangian
 (\ref{sine-Gordon}).
 Since the string world sheet is identified with the domain 
wall
in the effective action, we will integrate in the partition
function over all degrees of freedom apart from those that 
mark
the position of the domain wall. To do this let us split the 
field
$\phi$ into a continuous and discrete part
\begin{equation}
\phi(x)=\hat\phi(x)+2\pi\eta(x) \label{decomp}
\end{equation}
where the field $\hat\phi$ is continuous but bounded within 
the
vicinity of one ``vacuum''
\begin{equation}
-\pi< \hat\phi(x)<\pi
\end{equation}
and the field $\eta$ is integer valued. Clearly, whenever
$\eta(x)\ne 0$, the field $\phi$ is necessarily not in the
vicinity of the trivial vacuum $\phi=0$. Thus, the points in 
space
where $\eta$ does not vanish mark in a very real sense the
location of a domain wall between two adjacent vacua. The
partition function $Z$ is given by the  path integral 
\begin{eqnarray}
\int
{\cal{D}}\hat\phi {\cal{D}}\eta \, &&\, \exp\left(- \frac{g^2}{16 
\pi^2}
\int d^3x \left[\frac{1}{2} \partial_\mu(\hat \phi+2\pi\eta)
\partial^\mu (\hat\phi+2\pi\eta) \right.\right.
\nonumber\\[0.3cm]
 && +\left.\left.
 M^2_\gamma (1- \cos \hat \phi)
\right]
\right)\, .
\label{Z2} 
\end{eqnarray}
At weak coupling the field $\hat\phi$ can be
integrated out in the steepest descent approximation. This 
means
that the equations of motion for $\hat\phi$ have to be 
solved in
the presence of the external source $\eta$, keeping in mind 
that
$\hat\phi$ only takes values between $-\pi$ and $\pi$. Let 
us
assume for this solution that the surfaces along which 
$\eta$ does
not vanish are few and far between. Also, we will limit the
possible values of $\eta$ to $0,1,-1$ corresponding to
 the ``dilute gas'' approximation. In fact,
as we will see below, this exhausts all physically allowed 
values
of $\eta$. Then the qualitative structure of the solution 
for
$\hat \phi$ is clear. In the bulk, where $\eta$ vanishes, 
the
field $\hat\phi$ satisfies normal classical equations. When
crossing a surface $S$ on which $\eta=1$, the field 
$\hat\phi$
must jump by $2\pi$ in order to cancel the contribution of 
$\eta$
to the kinetic term, since otherwise the action is 
ultraviolet 
divergent.
Thus, the solution is that of a ``broken wall''. Far from 
the
surface the field $\hat\phi$ approaches its vacuum value
$\hat\phi=0$. Approaching $S$ it increases to $\pi$, then it 
jumps
to $-\pi$ on the other side of the surface, and again 
approaches
zero far from the surface. The profile of the solution for 
smooth
$S$ (in this case a plane) is depicted in Fig.~1. Since 
outside the
surface the field $\hat\phi$ solves classical equations of 
motion,
clearly the profile of $\hat\phi$ is precisely the profile 
of the
domain wall we discussed above. As noted above, the purpose 
of the
discontinuity across the surface S is to cancel the would be 
ultraviolet
divergent contribution of $\eta$ to the kinetic term in the
action. Thus, the action of our solution is precisely
\begin{equation}
S[\eta]=\sigma S=\sigma \int d^2 \tau \frac{\partial
X_\mu}{\partial\tau^\alpha} \frac{\partial X_\mu
}{\partial\tau_\alpha} \, ,
\end{equation}
where $\tau$ are the coordinates on the surface $S$, and 
$X_\mu$
are the coordinates in the three-dimensional space-time. 
This is
precisely the Nambu-Gotto action of a free string.

\begin{figure}
\hskip 4cm
\psfig{figure=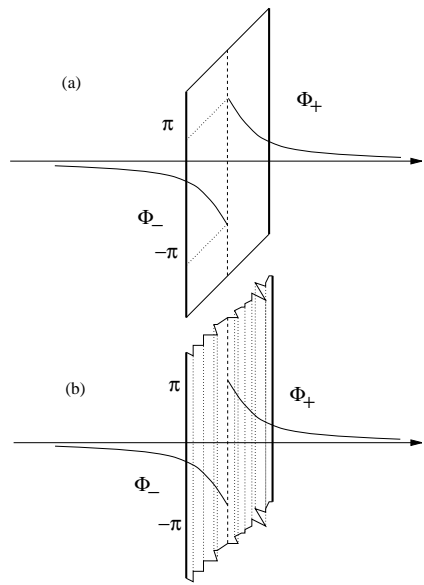,height=3.5in} \caption{Schematic
representation of the solution profile of the $\hat\phi$ 
field as
a function of the $x$-coordinate.} \label{fig:Fig.1}
\end{figure}

Of course the confining strings in the Georgi-Glashow model 
are
not free. The thickness of the region in Fig.~1 in which the 
field
$\hat\phi$ is different from zero, is clearly of order of 
the
inverse photon mass $M_\gamma^{-1}$. Thus, when two surfaces 
come
within distances of this order they start to interact. In
principle this interaction can be calculated by just finding 
the
classical solution for $\hat\phi$ in this situation. Now 
however
we want to discuss one particular property of the confining
strings  ---  their rigidity.

\subsection{The string is soft ... and therefore rigid!}

The Nambu-Gotto action we have derived in the previous 
subsection
is of course only the long wavelength approximation to the 
action
of the confining string. It is only valid for string 
world sheets
which are smooth on the scale of the inverse photon mass.
Expansion in powers of the derivatives can in principle be
performed and it will give corrections to the action on 
scales
comparable to $M^{-1}_\gamma$. However, for our confining 
string,
strange things happen in the ultraviolet. Physically, the
situation is quite peculiar, since the action of the string 
has
absolutely no sensitivity to changes of the world sheet on 
short
distance scales. This should be obvious from our derivation 
of the
string action. Suppose that, rather than taking $\eta=1$ on 
an
absolutely smooth surface, we make the surface look the same 
on
the scale $M_\gamma^{-1}$ but add to it some wiggles on a 
much
shorter distance scale $d$, as in Fig.~2. To calculate the 
action
we now have to solve the classical equation for the field
$\hat\phi$ with the new boundary condition  ---  the surface 
of the
discontinuity is wiggly. This new boundary condition changes 
the
profile of the classical solution only within the thickness 
$d$ of
the old surface. However, since the action of the classical
solution is in no way dominated by the region of space close 
to
the discontinuity, the action of the new solution will be 
the same
as that of the old solution with the accuracy $dM_\gamma$. 
Thus,
all string configurations which differ from each other only 
on
small resolution scales, $d\ll M^{-1}_\gamma$, have to this
accuracy the same energy! The string is therefore extremely 
soft,
in the sense that it tolerates any number of wiggles on 
short
distance scale without cost in energy, Fig.~3. In 
particular, since
the string tension for our string is much greater than the 
square
of the photon mass, $\sigma/M_\gamma^2=g^2/M_\gamma\gg 1$,
fluctuations on the scale of the string tension are not 
penalized
at all.

\begin{figure}
\hskip 4cm
\psfig{figure=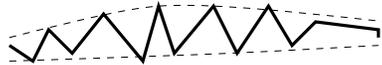,height=1.5in} \caption{The solid line
represents the actual thin string. The dashed lines denote 
the
contours of the effective thick string.} \label{fig:Fig.2}
\end{figure}

This independence of the action on short wavelength 
fluctuation is
a dynamical manifestation of the so called zigzag symmetry
introduced by Polyakov.\cite{POL96} Indeed, Polyakov notes 
that if
a segment of a string goes back on itself, physically 
nothing has
happened and so such a zigzag should not cost any action. 
This
situation is an extreme example of a wiggle we have just 
discussed
 ---  a wiggle with infinite momentum. What happens 
physically is that
not only do infinite momentum modes cost nothing, but so do 
finite
but large momentum string modes.

The confining string is therefore very different from a 
weakly
interacting string usually considered in the string theory. 
In the
weakly interacting string, momentum modes with momentum of 
order
of the square root of the string tension carry energy which 
is of
the same order. In the confining string on the other hand, 
these
momentum modes do not carry energy at all. Thus, we do not 
expect
the spectrum of the confining string to contain states with 
large
spatial momentum (heavy states with low angular momentum). 
In fact,
the states with high angular momentum may very well be 
absent too,
due to the fact that a long string can decay into a $W^+-W^-
$
pair. Therefore, it is likely that most of the string states 
will
not appear in the spectrum of the Georgi-Glashow model. 
Thus, our
confining string theory is quite peculiar: it is a string 
theory
without stringy states in the spectrum.

In a somewhat perverse way, this softness of the 
``mathematical''
string leads to rigidity of the physical string. The point 
is the
following. As is obvious from the previous discussion, the 
high
momentum modes are in a sense ``gauge'' degrees of freedom. 
The
action of any string configuration does not depend on them.
Consider a calculation of some physical quantity $O$ in the 
string
path integral. If $O$ itself does not depend on the string 
high
momentum modes, the integration over these modes factors 
out. If,
on the other hand, $O$ does depend on them, then its average
vanishes since their fluctuation in the path integral is
completely random. This is absolutely analogous to the 
situation
in gauge theories, where all observables must be gauge 
invariant,
and for calculation of those the integration over the gauge 
modes
always factors out, independently of the observable under
consideration.

Thus, for all practical purposes, we should just exclude the 
high
momentum string modes from consideration altogether. For 
example,
when calculating the entropy of our string, we should only 
take
into account the states which are different on the coarse 
grained
level with the coarse graining scale of order $M^{-
1}_\gamma$.
This means that our string is intrinsically ``thick''. If we 
still
want to describe this situation in terms of a mathematical 
string
with the string tension $\sigma$, we must make the string 
rigid so
that any bend on the scale smaller than $M_\gamma^{-1}$ is
suppressed. Thus, such a string theory must have a curvature 
term
with the coefficient of order $\sigma/M^2_\gamma$
\begin{equation}
\delta S\propto{\sigma\over M^2_\gamma}({\rm curvature}).
\end{equation}

 This is indeed what happens in the  truncated
confining string action (\ref{stringa}).

\subsection{The $W^\pm$ bosons  ---  the $0$ branes of 
the confining 
string}

Interestingly enough, the charged states of the 
Georgi-Glashow
model have a natural place in the confining string 
framework. To
represent a $W$ boson in the string language, let us 
consider a
path integral Eq.~(\ref{sine-Gordon}) in the presence of one
soliton.  As we discussed above, within the effective 
Lagrangian (\ref{sine-Gordon}) proper, solitonic 
configurations do 
not exist. However, they can be forced onto the system with 
the help of
the external current. To create a winding state with world 
line
$C$ we should insert a source in the path integral which 
forces a
unit winding on the field $\chi$. In other words, the field 
$\phi$
would have to change by $4\pi$ when going around $C$. The 
relevant
partition function is \bea Z[C]=\int D\phi\exp \left( - M_WL 
-
S[\phi, j]\right) \eea where \bea
 S[\phi, j] = \frac{g^2}{16\pi^2}
\int d^3x \left[\frac{1}{2} (\partial_\mu \phi- 2\pi j_\mu)
(\partial^\mu \phi-2\pi j_\mu) + M^2_\gamma (1-\cos 
\phi)\right].
\eea Here the ``external current'' $j_\mu$ is
\begin{equation}
j_\mu(x)=n^1_\mu(x)\delta(x\in S_1)+n^2_\mu(x)\delta(x\in 
S_2)
\label{surf}
\end{equation}
where $S_1$ and $S_2$ are two surfaces which both terminate 
on the
curve $C$, and the unit vectors $n^1$ and $n^2$ are the 
normal
vectors to these surfaces. The shape of the surfaces on 
which the
current $j_\mu$ does not vanish is illustrated in Fig.~4. 
The
insertion of $j_\mu$ forces the field $\phi$ to jump across 
the
surface $S_1$ by $2\pi$ and again jump by $2\pi$ across 
$S_2$ in
order to cancel the, otherwise ultraviolet divergent 
contribution of
$j_\mu$. Thus, when going around $C$ the field $\phi$ 
changes by
$4\pi$ and therefore $C$ is the world line of the $W^+$ 
boson. The
position of the surfaces $S_1$ and $S_2$ is arbitrary as 
long as
they both terminate on $C$. In particular, they could 
coincide, but
here we prefer to use a more general form with 
non-coinciding surfaces.

As before, splitting the field variable $\phi$ into 
$\hat\phi$,
and $\eta$ we see that the presence of $j_\mu$ just shifts 
the
variable $\eta$ by unity on the two surfaces $S_1$ and 
$S_2$. The
integration over $\hat\phi$ at fixed $\eta$ is performed in
exactly the same way as before. The only difference now is 
that
for any given $\eta$ one has two extra string world  sheets 
along
$S_1$ and $S_2$. After the integration over $\eta$, the 
result
will not depend on the exact position of $S_1$ and $S_2$. 
However,
at fixed $\eta$ the two surfaces introduced in 
Eq.~(\ref{surf})
specify the positions of the two extra world  sheets in the
confining string path integral. We thus see that the
field-theoretical path integral in the presence of a $W^+$ 
boson,
in the string representation, is given by the sum over 
surfaces in
the presence of a $0$  ---  brane, which serves as a source 
for two
extra string world  sheets.

Let us stress here the similarity and differences
 between $W$ particles viewed  as $0$-branes, and D0 branes 
in 
string theory.
Usually D0 branes are thought of as infinitely
heavy. The situation in the Georgi-Glashow model is very 
similar in this
respect. They are not infinitely heavy, but very heavy 
indeed
since the mass of $W$ is large on all scales relevant to 
zero
temperature physics.  However, this  analogy with D branes
 in superstring theory is not exact. The mass of a  D0 brane 
is  proportional to
$1/g_{str}$ whereas the mass of the $W$ boson (our $0$ 
brane) is
proportional to $g$. The fact that it is  very heavy in the  
weak
coupling limit is due to the large value of the Higgs 
condensate.
It may be more natural to think of the $W$ as of a stretched
F-string, which only after a duality transformation becomes 
a
D-object on which the confining string can end. For the sake 
of
brevity, in the following we will disregard this subtlety 
and will
continue to refer to the $W$ bosons as $0$ branes.

The contribution of the $0$ brane to the partition function 
is
suppressed by a very small factor $\exp\{-M_WL\}$, and 
vanishes
for an infinitely large system. As we will see in the later
sections however, the situation changes dramatically at 
finite
temperature, where one dimension of the system has finite 
extent.

Incidentally, going back to our definition of the vortex 
field
$V$, we see that in between the two world  sheets the value 
of $V$
is negative, while outside it is positive. Thus, we have 
created a
domain of the second vacuum of $V$ in between $S_1$ and 
$S_2$. It
may be easier to visualize the situation with both $W^+$ and 
$W^-$
present. In this case the surfaces $S_1$ and $S_2$ terminate 
on
one side on the world  line of $W^+$, and on the other side 
on the
world  line of $W^-$ (see Fig.~5). They are thus boundaries 
of a
closed domain of the second vacuum of the field $V$. In the
infrared therefore our strings are nothing but the Ising 
domain
walls, and the pair of D0 branes creates an Ising domain.

Note that in physical terms there are only two distinct 
vacua in
the model: $\langle V\rangle=1$ and $\langle V\rangle =-1$. 
Thus, having two domain 
walls is
the same as having a wall and an anti-wall, and if they 
coincide
spatially such a configuration is equivalent to the vacuum. 
A configuration of $\eta$ with $\eta=2$ on a closed surface 
is physically equivalent to vacuum. Therefore, the values 
that  $\eta$
is allowed to take are limited to $0$,\,\,\, $1$ and $-1$.

We close this section by noting, that the reason we have two
string world  sheets terminating on a D0 brane is that the 
field
theory in question only has fields in the adjoint 
representation
of SU$(2)$. We can imagine adding heavy fundamental 
particles to
the model. We would then also have allowed configurations of 
one
string world  sheet terminating on a D0 brane. These D0 
branes
would however be physically different and would have a 
different
mass and therefore a different weight in the path integral. 
In
field-theoretic terms, presence of the fundamental charges
drastically changes the properties of the $Z_2$ magnetic 
symmetry,
turning it into a local, rather than a global, 
symmetry.\cite{fosco}

\section{The Hamiltonian approach to compact U$(1)$}
 Our discussion so far has been in
the context of the Lagrangian path integral formalism. 
Often,
Hamiltonian approaches tend to be useful and more 
instructive. In
this section we will therefore show how the same confinement
phenomenon manifests itself in the canonical Hamiltonian 
analysis.

To make life a little easier in this section we will not 
study the
SU$(2)$ Georgi-Glashow model, but rather compact U$(1)$ 
theory.
This theory is also confining, and at low energies is
indistinguishable from the Georgi-Glashow model. Since we 
will not
at this point ask questions about the charged sector of the
theory, compact U$(1)$ is perfectly adequate for our 
purposes. In
this section we follow Refs. \citebk{var} and \citebk{ben}.

We start therefore by setting up the Hamiltonian description 
of
U$(1)$ compact QED. The first question to be settled, is 
what is
the Hilbert space of admissible states. Firstly, it is clear 
that
the Gauss' law should be implemented, and thus all the 
physical
states should satisfy
\begin{equation}
\exp\{i\int 
d^2x\partial_i\lambda(x)E_i(x)\}|\Psi\rangle=|\Psi\rangle.
\label{constr}
\end{equation}
There is a crucial difference between the Gauss' law in the
compact theory and in the noncompact one. In the noncompact 
theory
equation (\ref{constr}) should be satisfied only for regular
functions $\lambda$. For example, the operator
\begin{equation}
V(x)=\exp\left\{\frac{i}{g}\int d^2y\frac{\epsilon_{ij}(x-
y)_j}{(x-
y)^2}
E_i(y)\right\}
\label{vortex}
\end{equation}
which has the form of  (\ref{constr}) with the function 
$\lambda$
proportional to the planar angle $\theta$, i.e.
$\lambda=\frac{1}{g} \theta(x)$, does not act trivially on
physical states. In fact, this operator creates point-like
magnetic vortices with magnetic flux $2\pi/g$ just like
 Eq.~(\ref{v}). It therefore changes the physical state on 
which it
is acting.

In the compact theory the situation in this respect is quite
different. Point like vortices with quantized magnetic flux 
$2\pi
n/g$ cannot be detected by any measurement. In the Euclidean 
path
integral formalism of Ref. \citebk{polyakov} this is the 
statement
that the Dirac string of the monopole is unobservable and 
does not
cost any (Euclidean) energy. In the Hamiltonian formalism 
this
translates into the requirement that the creation operator 
of a
point-like vortex must be indistinguishable from the unit
operator. In other words, the operator (\ref{vortex}) 
generates a
transformation which belongs to the compact gauge group, and
should therefore act trivially on all physical states. 
Equation
(\ref{constr})
 should therefore
be satisfied also for these operators.

Accordingly, the Hamiltonian of the compact theory also must 
be
invariant under these transformations. The magnetic field, 
defined
as $B=\epsilon_{ij}\partial_iA_j$, on the other hand does  
not
commute with $V(x)$ (cf. Ref. \citebk{vort}),
\begin{equation}
V^\dagger (x)B(y)V(x)=B(y)+\frac{2\pi}{g}\delta^2(x-y).
\label{commut}
\end{equation}
The Hamiltonian should therefore contain not $B^2$ but 
rather a
periodic function of $B$. We will therefore choose our 
Hamiltonian
to be
\begin{equation}
H=\frac{1}{2}a^2\sum E^2_{\bn i}-\frac1{g^2a^2}\sum \cos
ga^2B_{\bn}. \label{ham}
\end{equation}
Since we will need in the following an explicit ultraviolet
regulator, we use lattice notations. Here $a$ is the lattice
spacing, and the sums are respectively over the links and
plaquettes of the two-dimensional spatial lattice. The
coefficients of the two terms in the Hamiltonian are chosen 
so
that in the weak coupling limit, upon formal expansion to 
lowest
order in $g^2$, the Hamiltonian reduces to the standard free
Hamiltonian of 2+1 dimensional electrodynamics. Following
Polyakov's book,\cite{book} we work in the weakly coupled 
regime. 
Since
the coupling constant $g^2$ in 2+1 dimensions has dimension 
of
mass, weak coupling means that the following dimensionless 
ratio
is small
\begin{equation}
g^2a\ll 1.
\end{equation}

Our aim now is to find the vacuum wave functional of this 
theory.
Exact solution of this problem is not feasible. However, at 
weak
coupling we can use the variational approximation.

\subsection{The variational ansatz}

For a weakly coupled theory one expects the vacuum wave 
functional
(VWF) to be not too different from the vacuum of a free 
theory.
Since the  VWF of free (noncompact) electrodynamics is 
Gaussian in
the field basis,
\begin{equation}
\psi_0[\bA]= \exp\left[-\frac12\sum_{\br,\bs}A_{\br i}
G^{-1}(\br-\bs)A_{\bs i}\right]\ , \label{wf0}
\end{equation}the
Gaussian variational approach in this case should give a 
good
approximation. An important caveat, however is that the 
ground
state WF should be gauge invariant under the full compact 
gauge
group. As a result it turns out that one can not take just a
Gaussian in $\bA$, since this will not preserve gauge 
invariance.
The simplest generalization of the Gaussian ansatz which we 
use
along the lines of Ref. \citebk{var}, is to project a 
Gaussian WF 
into
the gauge invariant subspace of the Hilbert space. The 
projection
has to be performed with respect to the full compact gauge 
group
of Eq.~(\ref{constr}).

To facilitate this, define a vortex field $A^V_{\bn i}$ that
satisfies (we suppress the lattice spacing $a$ henceforth)
\begin{equation}
\left(\nabla\times\bA^V\right)_{\bn'}=\frac{2\pi}g\delta_{{\
bn'},0}
\ ,\qquad \nabla\cdot\bA^V=0\ , \label{vortex1}
\end{equation}
where $\bn'$ is a plaquette on the lattice, or, 
equivalently, 
a site
of the dual lattice. This is the vector potential 
corresponding to
a magnetic field that is zero everywhere except at $\bn'=0$, 
where
it takes the value $\frac{2\pi}g$. The explicit solution of
(\ref{vortex1}) is
\begin{equation}
A^V_i(\bn)=-\frac{2\pi}g\epsilon_{ij}
\left(\frac{\partial_j}{\nabla^2}\delta_{\bn',0} 
\right)_{\bn}
\end{equation}
The compact gauge invariance requires that the variational 
wave
function $\psi[\bA]$ be invariant under shifts 
$\bA\to\bA+\bA^V$.
This of course is consistent with the periodicity of $H$ 
under
$B\to B+\frac{2\pi}g$. We also demand noncompact gauge 
invariance
of the wave function. Hence we define a field, shifted by a
noncompact gauge transformation $\phi_\bn$ and by a vortex
distribution $m_{\bn'}$,
\begin{equation}
\bA^{(\phi,m)}_\bn=\bA_\bn-(\nabla\phi)_\bn
-\sum_{\bn'}m_{\bn'}\bA^V(\bn-{\bn'}) \label{shift1}
\end{equation}
or, for short,
\begin{equation}
A^{(\phi,m)}=A-\nabla\phi-A^V\cdot m\ . \label{shift2}
\end{equation}
We choose the gauge invariant and periodic trial wave 
function as
\begin{equation}
\psi[\bA]=\sum_{\{m_{\bn'}\}}\int[d\phi_\bn]\,
\exp\left[-\frac12\sum_{\br,\bs}A^{(\phi,m)}_{\br i}
G^{-1}(\br-\bs)A^{(\phi,m)}_{\bs i}\right]\ . \label{wf}
\end{equation}
Under a gauge transformation,
\begin{equation}
\psi[\bA+\nabla\lambda]=\psi[\bA] \label{g_invariance}
\end{equation}
since $\lambda$ can be absorbed in a shift in $\phi$. The 
simple
rotational structure of $G_{ij}=\delta_{ij}G$ that appears 
in the
variational wave function (\ref{wf}) is consistent with
perturbation theory. We also take $G(x)$ to be a real 
function.

Our task now is to calculate the expectation value of the
Hamiltonian in this state, and to minimize it with respect 
to the
variational function $G$. We start by considering the
normalization of the wave function.

\subsection{Normalization integral}
The normalization of $|\psi\rangle$ is
\bea
Z & \equiv &\langle\psi|\psi\rangle = \sum_{\{m,m'\}}\int[d\phi][
d\phi'][d\bA]\nonumber\\[3mm]
&\times&
\exp\left[{-\frac12A^{(\phi,m)}G^{-1}A^{(\phi,m)}}\right]
\exp\left[{-\frac12A^{(\phi',m')}G^{-1}A^{(\phi',m')}}\right] 
.\label{norm1}
\eea
We shift \bA\ by $\nabla\phi'+\bA^V\cdot m'$, and absorb the 
shift
into $\phi$ and $m$, giving
\begin{equation}
Z=\sum_{\{m\}}\int[d\phi][d\bA]\,
e^{-\frac12A^{(\phi,m)}G^{-1}A^{(\phi,m)}} e^{-\frac12AG^{-
1}A}\ .
\label{Z}
\end{equation}
Now we combine the exponents according to
\bea
A^{(\phi,m)}G^{-1}A^{(\phi,m)}+AG^{-1}A &=& 2A^{(\phi/2,
m/2)}G^{-1}A^{(\phi/2, m/2)} 
\nonumber\\[0.3cm]
&+&\frac12 S(\phi,m)G^{-
1}S(\phi,m)\,,
\eea
where
\begin{equation}
{\bf S}\equiv{\bf \nabla}\phi+{\bf A}^V\cdot m\ .
\end{equation}
The first term in {\bf S} has zero curl while the second is
divergenceless. Furthermore, $G^{-1}$ is translation 
invariant and
proportional to the unit matrix.  Thus, $SG^{-1}S$ has no 
cross
terms between $m$ and $\phi$. We now shift \bA\ by
$\nabla\phi/2+\bA^V\cdot m/2$, and all the fields decouple. 
We
have then
\begin{equation}
Z=Z_A Z_{\phi} Z_v\ ,
\end{equation}
where
\begin{equation}
Z_A=\det \pi G\ ,
\end{equation}
\begin{eqnarray}
Z_{\phi} &=&\int[d\phi]e^{-\frac14\nabla\phi\cdot
G^{-1}\cdot\nabla\phi}\cr &=&\left(\det
4\pi\frac1{\nabla^2}G\right)^{1/2}\ ,
\end{eqnarray}
\begin{equation}
Z_v=\sum_{\{m_{\bn'}\}}\exp\left[-
\frac1{4g^2}\sum_{\br',\bs'}m_{\br'}
D({\br'}-{\bs'})m_{\bs'} \right]\ . \label{Zv1}
\end{equation}
Here $Z_v$ is the vortex partition function, with the 
vortex-vortex interaction $D$ given by
\begin{equation}
D({\br'}-{\bs'})=g^2\sum_{\br,\bs}\bA^V({\br}-{\br'})\cdot
G^{-1}(\br-\bs) \bA^V(\bs-\bs')\ ,
\end{equation}
or
\begin{equation}
D=-g^2A^VG^{-1}A^V=-4\pi^2\frac1{\nabla^2}G^{-1}\ . 
\label{DG1}
\end{equation}
We can split off the $\br'=\bs'$ terms in (\ref{Zv1}) and 
write
\begin{equation}
Z_v=\sum_{\{m_{\bn'}\}}\exp\left[-
\frac1{4g^2}\sum_{\br'\not=\bs'}m_{\br'}
D({\br'}-{\bs'})m_{\bs'} \right] \prod_{\br'}z^{m_{\br'}^2}
\label{Zv2}
\end{equation}
where we have defined the vortex fugacity
\begin{equation}
z=e^{-\frac1{4g^2}D(0)}\ . \label{fugacity1}
\end{equation}

In the interest of clarity, we adopt henceforth a continuum
notation, indicating the ultraviolet cutoff $a$ only where
necessary. Moreover, since the interesting physics comes 
from the
infrared, lattice effects can be approximated by a 
momentum-space
cutoff $\Lambda=a^{-1}$, which can simplify formulas 
further.

The variational function $G$ appears both in the 
vortex-vortex
potential and in the vortex fugacity. We expect the 
ultraviolet 
behavior of
$G$ to be the same as in the free theory, {\em viz.},
\begin{equation}
G^{-1}(k)\sim k\ ,
\end{equation}
so
\begin{eqnarray}
D(0)&\sim&\int^\Lambda\frac{d^2k}{(2\pi)^2}\frac{4\pi^2}{k^2
}G^{-
1}(k)
\label{DG}\\
&\sim&2\pi\Lambda
\end{eqnarray}
and thus
\begin{equation}
z\sim e^{-\frac{\pi}2\frac1{g^2a}} \label{potential}
\end{equation}
In the weak coupling region we have $z\ll1$, justifying a
restriction to $m=0,\pm1$ in  (\ref{Zv1}) and (\ref{Zv2}).

\subsection{Expectation values}

We calculate correlation functions of $m$ via a duality
transformation.\cite{Coleman} We add an $iJ\cdot m$ term to 
the
exponent in (\ref{Zv1}) and use the formula
\begin{equation}
e^{-\frac1{4g^2} m\cdot D\cdot m} ={\it const}\int[d\chi]\,
e^{-g^2\chi\cdot D^{-1}\cdot\chi} e^{i\chi\cdot m} 
\label{duality}
\end{equation}
to obtain
\begin{equation}
Z_v=\int[d\chi]\, e^{-g^2\chi\cdot D^{-1}\cdot\chi} 
\prod_{\bn}
[1+2\cos(\chi_{\bn}+J_{\bn})] .\label{sin1}
\end{equation}
Noting that\,\footnote{The normal ordering is performed 
relative to
the free theory defined by the quadratic action in 
(\ref{sin1}).}
\begin{equation}
\cos (\chi+J)=\eval{\cos\chi}_0:\!\cos 
(\chi+J)\!:\,=z:\!\cos
(\chi+J)\!: \label{normalorder}
\end{equation}
we have
\begin{eqnarray}
Z_v&=&\int[d\chi]\, e^{-g^2\chi\cdot D^{-1}\cdot\chi}
\prod \left[1+2z:\!\cos (\chi+J)\!:\right]\nonumber\\
&\simeq& \int D\chi\exp\left[-g^2\chi D^{-1} \chi+ 2z\int
d^2x\,:\!\cos\Big(\chi(x)+J(x)\Big)\!:\right]\ , 
\label{sine}
\end{eqnarray}
in continuum notation. Correspondingly,\cite{var}
\begin{equation}
\eval{m(x)m(y)}=2g^2D^{-1}(x-y)-4g^4\eval{D^{-
1}\chi(x)\,D^{-
1}\chi(y)}\
.
\end{equation}

The propagator of $\chi$ is easily calculated. To first 
order in
$z$, the only contribution comes from the tadpole diagrams, 
which
have already been subtracted in (\ref{sine}). Therefore,
\begin{eqnarray}
\int d^2x\,e^{ikx}\eval{\chi(x)\chi(0)}&=& \frac{1}{2g^2D^{-
1}(k)+2z}\nonumber\\[3mm]
&=&\frac{D(k)}{2g^2} -z\frac{D^2(k)}{2g^4} + O(z^2)\ .
\end{eqnarray}
The correlation function of the vortex density is then
\begin{equation}
K(k)\equiv\int d^2x\,e^{ikx}\eval{m(x)m(0)}=2z + O(z^2)\ ,
\label{rhoc}
\end{equation}
which in this approximation does not depend on momentum.

Now we are ready to calculate the expectation value of the
Hamiltonian (\ref{ham}). Using the definition (\ref{wf}) we 
obtain
\begin{eqnarray}
&&V^{-1}\eval{\int E^2\,d^2x}=-
\frac1V\eval{\psi\left|\sum_{\bn,i}
\frac{\partial^2}{\partial 
A_{\bn,i}^2}\right|\psi}\nonumber\\[0.3cm]
&&=\frac12\int \frac{d^2k}{(2\pi)^{2}} G^{-1}(k)-
\frac{\pi^2}{g^2}
\int \frac{d^2k}{(2\pi)^{2}}  k^{-2}G^{-
2}(k)K(k)\label{elen1}\\[3mm]
&&=\frac{1}{2}\int \frac{d^2k}{(2\pi)^{2}}
G^{-1}(k)-\frac{2\pi^2}{g^2} z\int \frac{d^2k}{(2\pi)^{2}}
k^{-2}G^{-2}(k) \label{elen}
\end{eqnarray}
The magnetic part is easily calculated since it has an 
exponential
form and, therefore, with our trial wave function leads to a 
simple
Gaussian integral. We find
\begin{equation}
\eval{e^{ingB_{\bn}}}=\exp\left[-{1\over 
4}n^2g^2\int{d^2k\over
(2\pi)^2}k^2G(k)\right]\eval{e^{in\pi m_{\bn}}}\ .
\end{equation}
The second factor is due to the vortices, and is different 
from
unity only for odd values of $n$. Using (\ref{sine}) we find
easily that $\eval{e^{i\pi m}}=e^{-4z}$. Expanding to 
leading
order in $g^2$ and $z$, we get
\begin{equation}
\eval{-\frac1{g^2}\cos gB} = \frac14\int {d^2k\over 
(2\pi)^2}
k^2G(k)+{4\over g^2}z\ , \label{Benergy}
\end{equation}
where we have dropped an additive constant. Finally, the 
expression
for the variational vacuum expectation value of the energy 
is
\begin{equation}
\frac{1}{V}\eval H= \frac{1}{4} \int\frac{ d^2k}{(2\pi)^2}
\left[G^{-1}(k)+k^2G(k) -\frac{4\pi^2}{g^2}
z\left(k^{-2}G^{-2}(k)- {4\over \pi^2}\right) \right]\ .
\label{eval}
\end{equation}

\subsection{Determination of the ground state}

The expression (\ref{eval}) has to be minimized with  
respect to
$G$.
 From equation (\ref{fugacity1}) and (\ref{DG})  we find
\begin{equation}
\frac{\delta z}{\delta G(k)}=\frac{1}{4g^2}k^{-2}G^{-
2}(k)\,z
\end{equation}
The variation of (\ref{eval}) gives\,\footnote{We have 
dropped  a
term $-\frac{8\pi^2}{g^2}zk^{-2}G^{-3}(k)$ from the 
right-hand
side of (\ref{variation})  is smaller by a factor of
$\frac{g^2k}{\Lambda^2}$ than the term retained, assuming 
$G\sim
k^{-1}$ at large $k$.}
\begin{equation}
{k^2-G^{-2}(k) = \frac{4\pi^4}{g^{4}}
z k^{-2} G^{-2}(k) \int \frac{d^2p}{(2\pi)^2}
\left[p^{-2}G^{-2}(p)- {4\over \pi^2} \right]}\ .
\label{variation}
\end{equation}
Equation (\ref{variation}) has the solution
\begin{equation}
G^{-2}(k)=\frac{k^4}{k^2+m^2} \label{solution}
\end{equation}
where
\begin{equation}
{m^2= \frac{4\pi^4}{g^4}
z\int \frac{d^2k}{(2\pi)^2} \left[k^{-2}G^{-2}(k)-
{4\over\pi^2}\right]} \ . \label{mass}
\end{equation}
The main contribution to the integral in the gap equation
(\ref{mass}) comes from momenta $k^2\gg m^2$. For these 
momenta
$k^2G^{-2}(k)=1$. Therefore, we see that (\ref{mass}) has a
nontrivial solution. Using Eqs. (\ref{fugacity1}),
(\ref{DG}), and (\ref{solution}) we obtain
\begin{equation}
{m^2= \frac{4\pi^4}{g^4}
\exp\left(-\frac{\pi^{2}}{g^{2}} \int \frac{d^2p}{(2\pi)^2}
\frac{1}{\sqrt{p^2 + m^{2}}}\right) \int 
\frac{d^2k}{(2\pi)^2}
\left[\frac{k^2}{k^2+m^2}-{4\over\pi^2}\right]}
\label{complicatedm}
\end{equation}
which for $g^2 a = g^2/\Lambda \ll1$ can be simplified to 
[cf.~(\ref{potential})]
\begin{equation}
{m^2 = {4\pi^2} \frac{(\pi^2-4)\Lambda^4}{g^4}
\exp\left(-\frac{\pi\Lambda }{2g^2}\right)}\ ,
\label{hamiltonianmassm}
\end{equation}
where we have restored the ultraviolet cutoff dependence
explicitly. The resulting $m$ is the mass gap of the theory, 
in
the sense that it is the inverse of the spatial correlation
length. Calculating, for example, the propagator of magnetic
field, we find
\begin{equation}
\eval{e^{igB_{\bm}}e^{-
igB_{\bn}}}=\left|\eval{e^{igB}}\right|^2
e^{\frac{g^2}2\nabla^2G(\bm-\bn)}\ ,
\end{equation}
and at large distances (neglecting power-like prefactors),
\begin{equation}
\nabla^2G(x)=-
\int\frac{d^2k}{(2\pi)^2}(k^2+m^2)^{1/2}e^{i\bk\cdot\bx}
\sim e^{-mx} \ .
\end{equation}
This dynamically generated mass is Polyakov's 
result.\cite{polyakov} 
Thus, we recover in the Hamiltonian approach the
first important result  ---   a finite mass gap $m$. Let us note
 that a mass gap $m$  in a Hamiltonian  formalism (\ref{hamiltonianmassm})
  has a  different preexponential factor in comparison with a mass gap
 $M_{\gamma}$ obtained in a path integral approach (\ref{Mgamma}). The reason
  is very simple. Contrary to universal exponential factor the 
preexponential factor  depends on details of UV regularizations which  are 
 different in  Lagrangian and hamiltonian formalisms.  
\subsection{Spatial Wilson loops}  

We also want to see whether the charges are confined in our 
best
variational state. The simplest quantity that is related to
confinement is the expectation value of the Wilson loop. 
Therefore, we 
will
calculate it in our ground state,
\begin{equation}
W_l[C]= \eval{\exp\left(ilg\oint_C \bA\cdot d\bx\right)} =
\eval{\exp\left(ilg\int_\Sigma B\,dS\right)}\ ,
\end{equation}
where $l$ is an integer and the integral is over the area 
$\Sigma$
bounded by the loop $C$. We have
\begin{equation}
W_l[C]=\eval{\prod_S e^{il\pi m_\bn}}Z_A^{-1}\int D\bA 
\exp\left(
-AG^{-1}A+ilg\int_\Sigma B\,dS\right) \label{wilson}
\end{equation}
The second factor is a Gaussian integral, which gives the 
factor
\begin{equation}
W_A=\exp\left(\frac{l^2g^2}4\int_\Sigma d^2x\int_\Sigma 
d^2y\,
\nabla^2G(\bx-\by)\right)\ . \label{wa}
\end{equation}
In the limit of large $\Sigma$ the leading behavior of the
exponent is
\begin{equation}
-\frac{l^2}4g^2\Sigma\lim_{k\to0}k^2G(k)=-
\frac{l^2}{4}g^2m\Sigma\
.
\end{equation}
This gives the area law with the string 
tension\,\footnote{As 
was
shown in Ref.~\citebk{greensite} the dependence on $l$ in 
this 
formula is
incorrect. The correct result is $\sigma\propto l$ rather 
than
$\sigma\propto l^2$. For multiply charged Wilson loops the
nonlinearities of the compact theory are important and the
Gaussian variational {\em ansatz\/} may be inadequate.}
\begin{equation}
\sigma=\frac{l^2}{4}g^2m\ . \label{str}
\end{equation}
The first factor in Eq. (\ref{wilson}) is different 
from
unity only for odd $l$. It can be easily calculated to be
\bea
W_v\equiv\eval{\prod_S e^{il\pi m_\bn}}&=& \int D\chi
\exp\left[-g^2\chi D^{-1}\chi\right.
\nonumber\\[0.3cm]
&+&2z\int d^2x\left. 
:\!\cos\Big(\chi(x)-\alpha(x)\Big)\!:\right]\ , 
\label{wvort}
\eea
where $\alpha(x)$ vanishes for $x$ outside the loop and is 
equal
to $\pi$ for $x$ inside the loop. At weak coupling we expand
around a classical minimum of the exponent. Recall that the
inverse propagator $D^{-1}$ is nonzero at zero momentum
Eq.~(\ref{DG1}). This dictates the leading order solution
$\chi(x)=0$ and
\begin{equation}
W_v=e^{-4z\Sigma}\ . \label{wvortf}
\end{equation}
This is a sub-leading correction to the string tension
(\ref{str}), since $g^2m\sim e^{-\pi/4g^2}$ and $z\propto 
m^2 \ll
g^2m$.

The behavior of the spatial Wilson loop suggests that the 
theory
is confining with the string tension related in the expected 
way
to the dynamically generated scale, $\sigma\propto g^2m$. 
However,
the spatial Wilson loop does not directly give the potential
between external charges. Although in the Euclidean 
formulation
there is no difference between spatial and time-like Wilson 
loops,
in the Hamiltonian approach one should be more careful. In
particular, the Hamiltonian formulation does not preserve 
the
Lorentz symmetry explicitly. It is therefore important to
calculate the potential between external charges directly.
 Our variational ansatz  can be extended
to calculate explicitly such a potential.\cite{ben}

\subsection{Interaction potential between external charges}
In the charged sector the gauge invariance condition
(\ref{g_invariance}) is replaced by
\begin{equation}
\psi[\bA+\nabla\lambda]=\psi[\bA] \exp\left(ig\sum_\bn
\rho_\bn\lambda_\bn\right)\ , \label{gauge3}
\end{equation}
where $\rho_\bn$ is a fixed, integer background charge
distribution. To calculate the interaction potential between 
two
charges we will later take it to be a well-separated dipole. 
A wave function that satisfies (\ref{gauge3}) is
\bea
\psi[\bA] &=& \sum_{\{m_{\bn'}\}}\int[d\phi_\bn]\,
\exp\left[-\frac12\sum_{\br,\bs}A^{(\phi,m)}_{\br i}
G^{-1}(\br-\bs)A^{(\phi,m)}_{\bs i}\right]\nonumber\\[3mm] 
&\times &\exp\left(ig\sum_\bn
\rho_\bn\phi_\bn\right)\ . \label{WF1}
\eea
The shifted field in (\ref{WF1}) is defined along the lines 
of
(\ref{shift2}),
\begin{equation}
A^{(\phi,m)}=A-\nabla\phi-A^S\cdot m\ . \label{as}
\end{equation}
but now it is necessary to define the vortex field $A^S$ in 
a
singular gauge. Like $A^V$, the field $A^S$ satisfies
\begin{equation}
\left(\nabla\times\bA^S\right)_{\bn'}=\frac{2\pi}g\delta_{{\
bn'},0}\  .
\end{equation}
While $A^V$ is divergenceless we take for $A^S$ the 
solution
where $A$ is non-zero on the links dual to a string 
extending from
0 (the plaquette of the vortex) to $x=+\infty$,
\begin{eqnarray}
A^S_{\bn x}&=&0 ,\cr A^S_{\bn y}&=&\left\{
\begin{array}{cl}
\frac{2\pi}g&{\rm for}\ n_x>0,\ n_y=0,\cr 0&{\rm otherwise}.
\end{array}
\right.
\end{eqnarray}

Our reason for using $A^S$ rather than $A^V$ in (\ref{WF1}) 
is one
of locality. A dipole can be created in the vacuum 
(\ref{wf}) by
the string operator 
$\exp\left[ig\sum_{\bx}^{\by}A(\bz)\right]$,
which places sources at \bx\ and \by. This operator creates 
a
string of electric flux taking integer values along the 
string,
the most local way of preserving Gauss' law when creating a
dipole. A shift of \bA\ by $A^S$ commutes with this string
operator, but a shift by $A^V$ does not. Shifting by $A^V$ 
will
create a non-local, transverse electric field with {\em
fractional} flux in addition to the string. In this light it
appears unavoidable that the introduction of dynamical 
charges
will immediately lead to a nonlocal and 
non-Lorentz-invariant
theory. For more on this point see Ref.~\citebk{DQSW}.

Another way of looking at it is to note that the difference
between $A^V$ and $A^S$ can be absorbed by a shift in the
integration variable $\phi_{\br}$ by ${1\over
g}\theta(\br-\br')m_{\br'}$, where $\theta(\br-\br')$ is the 
angle
that the vector $\br-\br'$ makes with the $x$ axis. Using 
$A^V$
would therefore lead to an extra phase factor
$$\exp\left[i\sum_{\br,\br'} \rho_\br
\theta(\br-\br')m_{\br'}\right]$$ in the integral in 
(\ref{WF1}).
Under a shift of vortex density $m$ this function is not
invariant, but rather acquires a phase proportional to the 
charge
density. The shift of vortex density can be viewed as a kind 
of
large gauge transformation\,\cite{var} and therefore this 
wave
function belongs to a sector of the Hilbert space with a
position-dependent ``$\theta$ angle.'' It is hard to imagine 
how
this sector can define a local theory, especially in the 
presence
of dynamical charges.

Since $A^S$ differs from $A^V$ by a gauge transformation, it 
can
be used interchangeably with it in the vacuum wave function
(\ref{wf}). The distinction is only meaningful in the 
charged
sector, where it leads to a new vortex-charge interaction 
(see
(\ref{part1}) below).

The wave function (\ref{WF1}) satisfies the Gauss' law, 
but it
turns out to be a very poor variational state.\cite{ben} The
reason is that it does not contain any extra variational 
parameter
on top of the width of the Gaussian. Thus, the way the 
electric
field, created by the external charges spreads in this state 
is
not optimized. One can in fact calculate the distribution of 
the
electric field.\cite{ben} This turns out to have a 
Coulomb-like
profile. As a result, the energy of such a state is infrared
divergent even for a finite size dipole. To do better we 
must
introduce an extra variational parameter which could 
optimize the
profile of the electric field. Thus, rather than using
Eq.~(\ref{WF1}) we take as out state
\begin{eqnarray}
\psi[\bA]&=&\sum_{\{m_{\bn'}\}}\int[d\phi_\bn]\,
\exp\left[-\frac12\sum_{\br,\bs}A^{(\phi,m)}_{\br i}
G^{-1}(\br-\bs)A^{(\phi,m)}_{\bs i}\right]
\nonumber\\[0.3cm]
&&\times \exp\left(ig\sum_\bn\rho_\bn\phi_\bn\right)
\exp\left(i\sum_\bn{\bf e}_\bn\cdot\bA^{(\phi,m)}
_\bn\right)\ . \label{wf2}
\end{eqnarray}
We take the classical background field ${\bf e}$ to be 
transverse,
$\nabla\cdot{\bf e}=0$, and we will treat it as an 
additional
variational parameter in the charged sector.

The normalization factor for the wave function (\ref{wf2}) 
is
\begin{equation}
Z=Z_AZ_\phi[\rho] Z_v [\rho]
\end{equation}
with
\begin{eqnarray}
Z_A&=&\det \pi G\ ,\\[3mm]
Z_\phi[\rho]&=& \left(\det 
4\pi\frac1{\nabla^2}G\right)^{1/2}
\exp\left(-g^2\rho\cdot\frac1{\nabla^2}G\cdot\rho\right)\ 
,\\[3mm]
Z_v[\rho]&=&\sum_{\{m_{\bn'}\}}\exp\left[-
\frac1{4g^2}\sum_{\br',\bs'}m_{\br'}
D({\br'}-{\bs'})m_{\bs'} \right]\nonumber\\[3mm]
&\times&
\exp\left(-i\sum_{\br'}h_{\br'}m_{\br'}\right)\ . 
\label{part1}
\end{eqnarray}
The new ingredient in (\ref{part1}) is a vortex-charge
interaction,
\begin{equation}
\sum_{\br'}h_{\br'}m_{\br'}\equiv
\sum_{\br\br'}[\rho_{\br}\theta(\br-
\br')+\frac{2\pi}g\frac1{\nabla^2}\epsilon_{ij}\partial_i(\b
r-\br')
e_j(\br)]m_{\br'}\ .
\end{equation}
The vortex interaction potential $D$ is again given by
Eq.~(\ref{DG1}).

Let us now calculate the expectation value of the electric 
field
in this state. Straightforward algebra gives
\begin{equation}
\eval{E_i}= g\frac{\partial_i}{\nabla^2}\rho- 
\frac{i\pi}{g}G^{-1}
\frac{\epsilon_{ij}\partial_j}{\nabla^2}\eval{m} +e_i\ .
\label{field1}
\end{equation}
 The energy of this state is
\begin{eqnarray}
\eval{H}&=&\frac V4\int\frac{d^2k}{(2\pi)^2}
\left[G^{-1}(k)+k^2G(k)\right]
-\frac{g^2}2\int d^2x\,\rho\frac1{\nabla^2}\rho\nonumber\\
&&\qquad +\frac{\pi^2}{2g^2}\int
d^2x\,d^2y\,\partial^{-2}G^{-2}(x-y)\eval{m(x)m(y)}
\nonumber\\&&\qquad -\frac{1}{g^2}\int d^2x\,\left({\rm
Re}\eval{e^{i\pi m(x)}}-1\right) \nonumber\\&&\qquad
+\frac{i}{2}\zeta G^{-1} \eval{m} +\frac12\int d^2x\,e^2\ ,
\label{energy2}
\end{eqnarray}
where the penultimate term contains the potential 
$\zeta(x)$,
defined via
\begin{equation}
e_i=\frac g{2\pi}\epsilon_{ij}\partial_j\zeta\ . 
\label{zeta}
\end{equation}

To calculate correlation functions of the vorticity $m$, we 
again
introduce the dual field $\chi$ as in Eq.~(\ref{duality}) 
and
obtain the relations
\begin{eqnarray}
\eval{m}&=&-2ig^2D^{-1}\eval{\chi},\nonumber\\[2mm]
\eval{m(x)m(y)}&=&2g^2D^{-1}-4g^4\eval{D^{-1}\chi(x)
D^{-1}\chi(y)}\ . \label{rhochi}
\end{eqnarray}
The duality transformation now results in the following 
Lagrangian
for the dual field $\chi$,
\begin{equation}
{\cal L}=g^2\chi D^{-1} \chi- 2z\int 
d^2x\,:\!\cos\Big(\chi(x)+\zeta
-\theta\rho\Big)\!:\ .
\end{equation}
The notation $\theta\rho$ represents the convolution of the 
source
distribution $\rho(\br)$ with $\theta(\br-\br')$.

To first order in $z$ we obtain
\begin{eqnarray}
\eval{m(x)}&=&-2iz\sin(\theta\rho(x)-
\zeta(x)),\nonumber\\[2mm]
\eval{m(x)m(y)}&=&2z\cos(\theta\rho(x)-\zeta(x))\delta^2(x-
y),
\label{rho2}\nonumber\\[2mm]
\eval{e^{i\pi
m(x)}}&=&\exp[-4z\cos(\theta\rho(x)-\zeta(x))]. 
\end{eqnarray}
The ${\bf e}$-dependent piece of the energy is
\begin{eqnarray}
\Delta E&=&m^2g^2\int d^2x\,[1-\cos(\theta\rho(x)-
\zeta(x))]\nonumber\\
&&\quad+z\int d^2x\,d^2y\,\zeta(x)
G^{-1}(x-y)\sin(\theta\rho(y)-\zeta(y))\nonumber\\
&&\quad-\frac{g^2}{8\pi^2}\int d^2x\,\zeta\nabla^2\zeta
.\label{en}
\end{eqnarray}
The quantity (\ref{en}) is to be minimized with respect to
$\zeta$. It is obvious immediately that at large distances 
from
the sources $\zeta\to\theta\rho$, so that the energy is 
infrared 
finite.
Noting that the second term is of order $g^2$ relative to 
the
first, we drop it from now on. The minimization equation for
$\zeta$ then becomes very simple,
\begin{equation}
{\nabla^2\zeta-m^2\sin(\zeta-\theta\rho)=0\ .} \label{sgmin}
\end{equation}

To study Eq.~(\ref{sgmin}) it is convenient to define
$\tilde\zeta=\zeta-\theta\rho$ which satisfies the 
sine-Gordon
equation with a singular source term,
\begin{equation}
{\nabla^2\tilde\zeta-m^2\sin\tilde\zeta=S} .\label{sineg}
\end{equation}
The source term $S$ consists of a dipole layer along the 
line
between the external point charges. For charges separated by 
a
distance $L$, we have
\begin{equation}
S(\bx)=2\pi\,\delta'(x_2)\,\eta\left(\frac L2+x_1\right)\,
\eta\left(\frac L2-x_1\right)\ ,
\end{equation}
where $\eta$ is a step function. When $L$ is much larger 
than
$1/gm$ the solution of Eq.~(\ref{sineg}) can be found in the 
region
$-L/2\ll x_1\ll L/2$. In this region $\tilde\zeta$ is
approximately independent of $x_1$ and satisfies the
one-dimensional sine-Gordon equation in $x_2$,
\begin{equation}
{\frac{\partial^2\tilde\zeta}{\partial
x^2_2}-m^2\sin\tilde\zeta=S\ .}
\end{equation}
The singular source $S$ creates a discontinuity of $2\pi$ at
$x_2=0$. The solution for $x_2>0$ is hence 
half a sine-Gordon
soliton, with $\tilde\zeta=\pi$ at $x_2=0+$ and 
$\tilde\zeta\to0$
as $x_2\rightarrow\infty$. For $x_2<0$ it is half 
the anti-soliton
with $\tilde\zeta=-\pi$ at $x_2=0-$ and $\tilde\zeta\to0$ as
$x_2\to-\infty$. At distances greater than $1/m$ from the 
$x_1$
axis, $\tilde\zeta$ vanishes exponentially. Therefore, we 
have
$\zeta\to\theta\rho$  exponentially. The vortex density 
$m(x)$
according to Eq.~(\ref{rho2}) vanishes exponentially. 
Referring to
Eq.~(\ref{field1}) we see that the total electric field 
vanishes
exponentially outside the region of width $1/m$, indeed 
forming a
flux tube of thickness $1/m$. This is perfectly in accord 
with our
expectation for the field profile in a confining theory.

In Eq.~(\ref{en}) we see that the energy density of the flux 
tube
is proportional to the energy of the sine-Gordon soliton 
solution
with the proportionality coefficient $g^2\over 4\pi^2$. The 
string
tension is therefore
\begin{equation}
\sigma={2\over\pi^2}\,g^2m\ .
\end{equation}
This is consistent with the result of the calculation of the
spatial Wilson loop in the previous subsection. We have thus
established that the electric field in the vacuum of compact 
QED
is squeezed into flux tubes, and that the energy of a large 
dipole
is proportional to its length.

\subsection{Relation between the Lagrangian and the 
Hamiltonian 
pictures}
Some words should be said about the relation of the
 two-dimensional Hamiltonian calculation
 to the three-dimensional Euclidean path integral. The
vacuum wave functional of the theory can be represented in 
path
integral formalism. To get the vacuum WF $\Psi[A]$ one 
should
calculate the path integral over the fields $A(x,t)$, with 
$t$
varying from $-\infty$ to $0$, with the boundary condition 
$A(x,
t=0)=A(x)$. To be more precise, in calculating the VEV of 
some
operator $O(t=0)$, one should split the time coordinate of 
the
plane with the time coordinate of the operator, so that one
considers $\Psi[A(x,t=-\epsilon)]$ and 
$\Psi^*[A(x,t=\epsilon)]$
in the limit $\epsilon\rightarrow 0$.

The basic objects that appear in the Euclidean path 
integrals are
monopoles, which in 3D are not propagating particles, but 
rather
instantons. When described in terms of the vector potential, 
or
noncompact field strength, a monopole has a Dirac string 
attached
to it. It is clear that the vortices (anti-vortices) in 
Eq.
(\ref{wf}) correspond precisely to the intersections of the 
Dirac
strings of the 3D monopoles (anti-monopoles) with the equal 
time
plane at $t=0$. The positions of the Dirac strings are not
physical in the compact theory, and only the position of the
monopole itself is gauge invariant. In fact, for all 
monopoles
that do not sit in the infinitesimally thin time slab 
between the
planes $t=-\epsilon$ and $t=\epsilon$, one can always choose 
the
direction of the Dirac string such that it does not 
intersect the
two planes. This precisely corresponds to the shift of the
integration variable in going from Eq.~(\ref{norm1}) to
Eq.~(\ref{Z}). The combination that enters this path 
integral
nontrivially is
 the difference in vorticity between $\Psi$ and $\Psi^*$. At 
points
where both $m$ (which corresponds to $\Psi$) and $m'$ (which
corresponds to $\Psi^*$) are nonzero but equal, the 
difference
vanishes. This is the situation when a Dirac string 
intersects
both planes $t=\pm\epsilon$. When a 3D monopole sits in the 
slab,
only one of $m$ or $m'$ does not vanish  and so their 
difference
is nonzero. The summation over $m$ in Eq. (\ref{Z})
 can be interpreted
therefore as the direct contribution to the partition 
function due
to the monopoles at precisely the time $t=0$.

The fact that in this way one sees directly only the 
monopoles at
$t=0$, does not mean of course that other monopoles are not 
taken
into account in this approximation. Indeed, the ``bare''
interaction potential between the $t=0$ monopoles is $D(x)$. 
In
the best variational state it is already short range, as 
follows
from the solution for $G(k)$ (see Eq. 
(\ref{solution})). This
is in accord with the 3D picture, where the 3D monopole gas
produces screening. Obviously, if one only looks at the thin 
slab,
every monopole there will have an anti-monopole partner, 
which
sits nearby (inside the screening length) in the third 
direction.
The 2D monopole gas will therefore be screened by the 3D
interaction, even before the interactions of the 2D 
monopoles
between themselves is taken into account. This is perfectly
consistent with our calculation. It is interesting to note, 
that
even though this 2D interaction produces additional 
screening (the
cosine term in the effective theory  (\ref{sine})),
 it is the 3D screening that is responsible for the
area law of the Wilson loop, as is clear from the 
calculation of
the string tension in  Eq.~(\ref{wilson}). In fact, if one 
takes
the noncompact expression $G^{-1}(k)=|k|$, even the 
subleading
term in the area law in the Wilson loop (\ref{wvortf})
disappears.\cite{var}

\section{The deconfining phase transition}

Now that we understand confinement in some detail, it is 
time to
ask how it disappears. The deconfining phase transition in 
QCD at
finite temperature has been the subject of numerous studies 
in
recent years. Some aspects of the physics of the high 
temperature
phase appear to be perturbative and can be studied in a 
controlled
way at asymptotically high temperatures. However, in the 
phase
transition region itself the QCD coupling is large, and the
physics is dominated by the nonperturbative soft sector. 
Analytic
understanding of this region is therefore extremely 
difficult. It
is thus again useful to fall back on our toy model where a 
similar
deconfining phase transition is within the weak coupling 
regime
and can be studied analytically.

In this section we discuss in detail the physics of the
deconfining phase transition.
 Analysis of the deconfining phase
transition in the compact U$(1)$ theory was performed in 
Ref.
\citebk{zarembo}. The SU$(2)$ Georgi-Glashow model was 
studied in
Ref.
\citebk{dunne}. Interestingly enough, although at zero 
temperature
the differences between these two models are minute, the 
behavior
close to the phase transition is very distinct. So much so 
that
the dynamics that drives the phase transition as well as the
universal characteristics of the transition are completely
different. In particular, whereas in the compact U$(1)$ 
theory the
transition is apparently driven by the binding of monopoles, 
in
the Georgi-Glashow model it is due to the plasma effects of
charged excitations ($W$ bosons). In the lattice framework 
the
monopole binding in the compact U$(1)$ theory has been 
recently
studied in Ref. \citebk{chernodub}. One word of caution 
though, is 
that
the compact U$(1)$ theory does not have  a proper continuum 
limit.
Thus, it is not entirely clear whether the monopole binding
mechanism is relevant to continuum field theory. In this 
review
therefore we will concentrate on the Georgi-Glashow model.

\subsection{The monopole binding}

Since at zero temperature the monopole contributions are the 
only
relevant ones for confinement, one may be tempted to assume 
that
also at finite, low enough temperature all other effects are
unimportant. The expected value of the transition 
temperature, as
we shall see in a moment, is of order $g^2$, which is much 
smaller
than any mass scale in the theory except for the photon 
mass. One
could therefore start with the working hypothesis that the 
single
self-interacting photon field (or, equivalently, the 
monopole
ensemble) should be a valid description of the phase 
transition
region. This  scenario was discussed in Ref. 
\citebk{zarembo}.

The physics here is simple. The first thing to note is that 
at
finite temperature the interaction between monopoles is
logarithmic at large distances. The reason is that the 
finite
temperature path integral is formulated with periodic 
boundary
conditions in the Euclidean time direction. The field lines 
are
therefore prevented from crossing the boundary in this 
direction.
The magnetic field lines emanating from a monopole have to 
bend
close to the boundary and go parallel to it. So effectively 
the
whole magnetic flux is squeezed into two dimensions. 
Qualitatively
the situation is shown in Fig.~3.

\begin{center}
\begin{figure}
\hskip 4cm
\psfig{figure=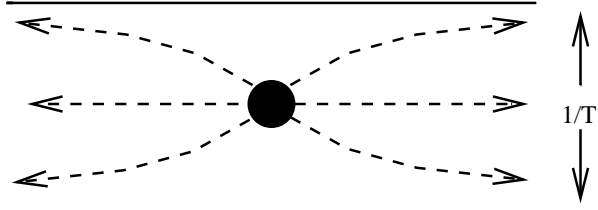,height=1.5in}
\caption{The field of a monopole-instanton at finite 
temperature.}
\label{fig:Fig.3}
\end{figure}
\end{center}

The length of the time direction is $\beta=1/T$, and thus 
the
field profile is clearly two-dimensional on distance scales 
larger
than $\beta$. Two monopoles separated by a distance larger 
than
$\beta$ therefore interact via a two-dimensional rather than 
a
three-dimensional Coulomb potential, which is logarithmic at 
large
distances. Since the density of the monopoles is tiny
$\rho_M\propto\xi$, the monopole gas becomes two-dimensional
already at extremely low temperatures $T\propto \xi^{1/2}$. 
The
strength of the logarithmic interaction is easily 
calculated. The
magnetic flux of the monopole should spread evenly in the 
compact
direction once we are far enough from the monopole core. The 
field
strength should have only components parallel to the spatial
directions. Since the total flux of the monopole is 
$2\pi/g$, the
field strength far from the core is $$\tilde F_i={T\over
g}{x_i\over x^2}\,,$$ and thus the strength of the infrared
logarithmic interaction is $T^2/g^2$.

It is well known\,\cite{BKT} that the two-dimensional 
Coulomb gas
undergoes the Berezinsky-Kosterlitz-Thouless
(BKT) phase transition. In the usual Coulomb 
gas,
where the strength of the interaction $\lambda$ does not 
depend on
temperature, the particles are bound in pairs at low 
temperature
and unbind at high temperature $T>T_{\rm BKT}=2\pi\lambda $, 
where the
entropy overcomes the energy. This is the standard BKT phase
transition\,\cite{BKT} which determines the universality 
class of
U$(1)$ symmetry restoration in 2 dimensions. In the present 
case
the situation is reversed, since the strength of the 
interaction
itself depends on the temperature. At low temperature the
interaction is weak, and therefore the particles (monopoles) 
are
free. As the temperature grows, the interaction becomes 
stronger
until at $T_{\rm BKT}=2\pi {T^2_{\rm BKT}/ g^2}$ the 
energy  overcomes
the entropy. Above this temperature, the monopoles bind into
neutral pairs. Thus, based on this simple picture one can 
expect
the theory to undergo a BKT phase transition at \be
T_{\rm BKT}={g^2\over 2\pi}. \label{TBKT} \ee Below this 
temperature
the photon should be massive, while above this temperature 
it
should be massless, since the cosine term in the Lagrangian
(\ref{sine-Gordon}) is irrelevant.

This is precisely the logic followed in Ref. 
\citebk{zarembo}. One 
can go
further and perform more quantitative calculations using the
dimensionally reduced version of Polyakov's effective 
Lagrangian (\ref{sine-Gordon}). Dimensional reduction should 
be 
perfectly
valid in this theory since the critical temperature is much 
larger
than the photon mass. The effect of the monopoles at finite
temperature is thus contained in the two-dimensional model
 \be {\cal L} = {g^2\over32 \pi^2 T}(\d_{i}\phi)^2
+ {M^2 g^2\over 16\pi^2 T}\cos \phi. \ee This theory, in 
full
agreement with the previous discussion has a BKT phase 
transition
at the temperature given by Eq.~(\ref{TBKT}), above which 
the
``dual photon'' field $\phi$ becomes massless.

Moreover, since the two-dimensional sine-Gordon theory is 
exactly solvable,
one can calculate various correlation functions. In 
particular, the
string tension was calculated in Ref. \citebk{zarembo}, and 
was 
found to
exhibits quite strange behavior. At low temperatures it 
follows an
expected pattern  ---  that is, it decreases with 
temperature. It
becomes extremely small at temperature much smaller than
$T_{\rm BKT}$; however, just before $T_{\rm BKT}$ it starts 
rising  sharply
and at $T_{\rm BKT}$ actually diverges.

Although the logic leading to the previous discussion seems 
quite
sound, there are several puzzles that arise regarding the 
results.

$\bullet$ The photon becomes massless in the high 
temperature phase. 
In
other words, the correlation length in some physical (gauge 
invariant)
channel becomes infinite at high temperature. This
contradicts our physical intuition, since one expects that 
at high
temperature any physical system described by a local field 
theory
should  become maximally disordered with zero correlation 
length.

$\bullet$ The phase transition is of BKT type, and therefore
belongs to the U$(1)$ universality class. On the other hand, 
we
know\,\cite{Kovner1}  that the global symmetry that is 
restored at 
the phase
transition is the magnetic $Z_2$, and we expect the
universality class to be 2D Ising. This is also expected 
from the
classic universality arguments.\cite{yaffe}

$\bullet$ The divergence of the string tension at the 
critical point 
looks
unphysical.

$\bullet$ Finally, as we explained in the previous sections, 
the
effective theory (\ref{sine-Gordon}) does not allow 
charged
states. It is then difficult to understand in what sense the 
high
temperature phase can be viewed as deconfined.

\subsection{Magnetic symmetry restoration and the charged 
plasma}
Another line of argument starts with the realization that
confinement is tantamount to the spontaneous breaking of the 
$Z_2$
magnetic symmetry. Indeed, let us now take Eq.
(\ref{lowlagrangian}) as our starting point. What does one 
expect
from the phase transition in a simple scalar theory of this 
type?
Firstly, clearly one expects the critical temperature to be 
of
order of the expectation value of the scalar field, and 
therefore
parametrically of order $g^2$. At finite temperature one 
expects
the generation of a positive thermal mass proportional to 
the
temperature and to the coupling constant. Thus, the thermal
contribution to the effective potential should be of the 
form \be
\delta_T
{\cal L} =x\lambda TV^*V \,.\ee At $T=g^2/4\pi^2x$ the total 
mass
term becomes positive, the VEV of $V$ vanishes and the phase
transition occurs. One can  estimate this temperature more
precisely using the following simple argument. Let us 
neglect for
now the monopole induced term. Then we are dealing with the 
$XY$
model at finite temperature. The dimensional reduction 
should
again be a valid approximation, and thus essentially we have 
to
analyze a 2D model. Now a 2D $XY$ model can be mapped into a
sine-Gordon theory of a dual field $\tilde\chi$. Performing 
this
dual transformation\footnote{To fix the normalization of the
kinetic term we should bear in mind that a vortex 
corresponds to a
$2\pi$ jump in the field $\chi=2\phi$ rather than the field 
$\phi$
in Eq.~(\ref{sine-Gordon}).}, we find the Lagrangian 
\be 
{\cal L} ={T\over
2g^2}(\partial_i\tilde\chi)^2+\mu\cos\tilde\chi \,.
\label{dualsG} \ee
where $\mu$ is the fugacity of the vortices in the $XY$ 
model.
This sine-Gordon theory has a phase transition at \be
T_{XY}={g^2\over 8\pi}. \ee Therefore, we may expect the 
magnetic
symmetry restoring phase transition at $T_{XY}$. Although
parametrically this temperature is of the same order as the 
BKT
transition temperature discussed in the previous subsection, 
it is
four times lower. Another important difference is that the 
nature
of the phase transition is in fact completely different. The 
phase
transition in the model (\ref{dualsG}) is due to the 
unbinding
of vortices of the field $V$. Above $T_{XY}$ the vortices 
are in
the plasma phase. These vortices are, as discussed earlier,
precisely the charged $W^\pm$ bosons of the original
Georgi-Glashow model. Thus, this phase transition is just 
what one
would naturally call the deconfining phase transition and 
the
vacuum above $T_{XY}$ is the charged plasma.

Note that in this discussion we have neglected completely 
the
effect of monopoles, that is the last term in the Lagrangian 
(\ref{lowlagrangian}). Interestingly enough we see that the 
plasma
phase is reached at a temperature which is much lower than
$T_{\rm BKT}$ discussed earlier, and thus the monopole 
binding 
is
irrelevant for the dynamics of the deconfinement. This is 
not to
say that the presence of the monopoles is irrelevant 
altogether.
Clearly, omitting the monopole induced term, we enlarged the
symmetry of the system from $Z_2$ to U$(1)$. Hence the 
effective
$XY$ model description and the U$(1)$ universality class 
predicted
by Eq.~(\ref{dualsG}). The analysis as it stands is correct 
for
noncompact electrodynamics with charged matter, but not for 
the
Georgi-Glashow model. Next we remedy this problem.

The effect of the monopoles can be qualitatively understood 
in a
simple way. Their presence leads to a confining potential 
between 
the
charges, that is a linear interaction between the $XY$ model 
vortices.
Thus, the monopoles suppress the variation of the phase of 
the field 
$V$
except inside the confining string, the width of which is 
given by 
the
inverse photon mass $d=1/M\propto\xi^{1/2}$. Whether the
presence of the monopoles is important at the would be
phase transition point $T_{XY}$
depends crucially on the density of charged particles. If 
the 
density
of the charged particles is very high, so that the average 
distance
between them is smaller than $d$, the presence of the
monopoles is immaterial since
the phase of $V$ is disordered already on the short distance
scale. However, if the density of charges is low, so that 
the 
distance
between them is larger than $d$, the presence of the 
monopoles will
suppress the phase transition. In this case at $T_{XY}$ the 
vacuum 
will
not be disordered, but will look like a dilute gas of 
charged 
particles
with strings between them. The actual phase transition will 
then 
occur
at a higher temperature, where the average distance between 
the
charges equals the inverse photon mass.

Interestingly enough in the present case this temperature 
turns
out to be twice the value of $T_{XY}$ and therefore still 
much
lower than $T_{\rm BKT}$. We will show this rigorously in 
the  next
subsection. Qualitatively though this is easy to understand. 
The
density of charges is proportional to their fugacity \be
\mu\propto e^{-M_W\over T}. \ee This should be compared to 
the
square of the photon mass, which is given by the zero 
temperature
monopole fugacity. In a theory with very heavy Higgs \be
\xi\propto e^{-4\pi M_W\over g^2}. \ee The two are equal (up 
to
subleading corrections) at \be T_{\rm GG}={g^2\over 4\pi}. 
\ee 
More
generally, when the Higgs is not infinitely heavy, the 
monopole
fugacity is smaller and thus the transition temperature will 
be
lower, \be T_c={g^2\over 4\pi\epsilon({M_H\over M_W})}. \ee 
For a
lighter Higgs, $T_c$ gets closer to $T_{XY}$, but is always
greater, so that $ T_{XY} < T_c < T_{\rm BKT}$.

In the next subsection we will present a more complete 
analysis
based on the renormalization group and exact bosonization. 
This
analysis confirms the simple picture presented here. Thus, 
all the
puzzles raised in the previous subsection simply disappear:

$\bullet$ The photon never becomes massless.
Even without the monopoles in the plasma phase it acquires 
the Debye
mass given by the cosine term in Eq. (\ref{dualsG}). This 
mass {\it
rises} with temperature, and thus the physical correlation 
length
decreases.

$\bullet$ Since the monopole term is still relevant at the 
phase
transition, the universality class must be $Z_2$. We will 
see this
explicitly in the next section.

$\bullet$ The analysis of the previous subsection is only 
valid 
below
$T_{XY}$, and thus the divergence of the string tension
at $T_{\rm BKT}>T_{XY}$ has nothing to do with the physics 
of 
the Georgi-Glashow model.

$\bullet$ Finally,
the phase transition is driven primarily by the unbinding of 
charged
particles and thus indeed has all the flavor of a 
deconfinement
transition.

Let us now turn to a more quantitative analysis of what is 
stated 
above.

\subsection{The  renormalization group analysis}

For a more formal analysis of the phase transition we find 
it
convenient to use the sine-Gordon formulation in terms of 
the
phase field $\chi$ modified to take into account explicitly 
the
finite probability of the appearance of vortices. This 
Lagrangian
has the form \be {\cal L} = { g^2 \over 8 \pi^2 T } 
(\d_{\mu}\chi)^2 
+
\zeta\cos 2 \chi + \mu \cos \tilde{\chi}\,, 
\label{sinevortex} 
\ee
where $\zeta$ is related to the monopole fugacity 
$\zeta=\xi/T$
and $\tilde \chi$ is the field dual to $\chi$, \be
i\d_{\mu}\tilde\chi= {g^2\over 2\pi T} 
\epsilon_{\mu\nu}\d^\nu
\chi\, . \label{chid} \ee The way to derive this Lagrangian 
is 
as
follows. The partition function of the sine-Gordon model in 
the
presence of one vortex is 
\be Z(x)=\int D[\chi]\exp\left[-\int 
d^2y {
g^2 \over 8 \pi^2 T } (\d_{\mu}\chi- j_\mu(y,x))^2 + \zeta 
\cos 2
\chi \right]\, .
\label{onev}
 \ee
 The ``external current'' is 
\be
j_\mu(y,x)=2\pi n_\mu(y)\delta(y\in C) \ee with $C$ a curve 
that
starts at the location of the vortex (the point $x$), and 
goes to
infinity, and $n_\mu$ is the unit normal to this curve. The
insertion of this current forces the derivative of $\chi$ to 
have
a discontinuity across the curve $C$, so that $\chi$ jumps 
by
$2\pi$. This forces $\chi$ to have one unit of vorticity
concentrated at the point $x$. Note that even though $j_\mu$
explicitly depends on the curve $C$, the partition function 
itself
does not, since changing the integration variable \be
\chi(x)\rightarrow\chi(x)+2\pi , \ \ \ \ \ \ \ x\in S \,,\ee 
where
the boundary of $S$ is $C-C'$, is equivalent to changing $C$ 
into
$C'$ in the definition of the current. The extra linear term 
in
the exponential in Eq.~(\ref{onev}) is \be i\tilde\chi= { 
g^2 \over
2 \pi T }\int_C dx_\mu\epsilon_{\mu\nu}\partial_\nu\chi\,,
\ee 
which
is equivalent to Eq.~(\ref{chid}). An anti-vortex at $y$ is
obviously created by $-j_\mu$. To create several vortices 
one just
inserts an external current which is the sum of the currents 
which
create individual vortices.

A dilute ensemble of vortices and anti-vortices with (small)
fugacity $\mu$ is then given by \be Z=\sum_{n,m}{1\over 
n!}{1\over
m!}\mu^{n+m}\int \Pi_idx_i\Pi_jdy_j Z(x_i,y_j) .\ee The 
summation
over the number of vortices and anti-vortices can be easily
performed leading to the partition function with the 
Lagrangian
(\ref{sinevortex}). The constant $\mu$ is the vortex 
fugacity
scaled by the effective ultraviolet cutoff imposed on the 
integration 
over
the coordinates. The vortex fugacity of course is none other 
but
the fugacity of the charged $W$, \be \mu=a^{-2} e^{-M_W\over 
T}\, .
\ee The cutoff $a$ is related to the Compton wavelength of 
the $W$
boson, but a more careful determination of it should take 
into
account the fact that in the process of dimensional 
reduction, all
modes with frequencies above $T$ have been integrated out. 
We will
not attempt the determination of $a$, but only note that it 
is
some combination of the scales $M_W$ and $T\propto g^2$, and 
as
such its value always plays a role secondary to the 
exponential
factor of fugacity.

An alternative way to derive the Lagrangian  
(\ref{sinevortex})
is to start directly from the effective theory 
(\ref{lowlagrangian}) 
with the extra Skyrme term  (\ref{skyrme}). 
In the nonlinear $\sigma$-model limit one can
cleanly separate the phase of the field $V$ into a smooth 
part
$\chi$  and the vortex contribution. The Skyrme term then is
proportional to the energy of one vortex and, with the 
Skyrme
coupling chosen the way we discussed in Sec. 2, is just 
equal
to $M_W$ for a one vortex configuration. The dilute vortex 
gas
approximation then reproduces the Lagrangian 
(\ref{sinevortex}).

 Since both $\xi$ and $\mu$ are small, the
importance of different terms in the Lagrangian 
(\ref{sinevortex}) 
is determined by their respective conformal
dimensions calculated in the free theory. The total 
conformal
dimensions of the operators $:\cos 2 \chi:$  and  $:\cos
\tilde{\chi}:$, respectively, are 
\be \Delta_\xi= {4\pi 
T\over g^2}
\hskip 1 cm {\rm and} \hskip 1 cm\Delta_\mu=  {g^2 \over 
4\pi T}\, .
\ee
 Thus, at low temperature  the charges are irrelevant
($\Delta_\mu>2$), the monopoles are relevant 
($\Delta_\xi<2$) and
the theory reduces to the sine-Gordon model. At the 
temperature
$T=g^2/4\pi$ the two operators become equally relevant, 
since
their conformal dimensions are equal, 
\be \Delta_\mu= 
\Delta_\xi=
1. \ee

The phase transition point can be determined from the 
structure of
the fixed points of the renormalization group equations. The
critical point is the infrared unstable fixed point of the 
renormalization group 
flow. The
renormalization group equations of the sine-Gordon theory 
have
been studied perturbatively in the literature both in the 
absence\,\cite{Coleman} and in the presence\,\cite{Kadanoff}  
of the 
vortices. In terms of the dimensionless 
parameters 
\be
t={4\pi\over g^2}T, \ \ \ \ \ \tilde\mu=\mu a^2, \ \ \ \ \ \
\tilde\zeta=\zeta a^2 
\ee 
the lowest order renormalization group equations read 
\bea
{dt\over d\lambda} & = &\pi^2(\tilde\mu^2 - t^2 
\tilde\zeta^2), \\
{d\tilde\mu\over d\lambda} & = & (2- {1\over t})\,\, 
\tilde\mu, \\
{d\tilde\zeta\over d\lambda} & = & (2-   t)\,\, 
\tilde\zeta\,.
\label{RG}
\eea
The fixed point structure of these equations is simple.

1. The point $T_0$
\be
t=0,\ \ \ \ \tilde\mu=0, \ \ \ \ \tilde\zeta=\infty
\ee
is clearly the zero temperature fixed point. Here the long 
distance
physics is dominated by the monopole induced mass term.

2. The point $T_\infty$
\be
t=\infty, \ \ \ \ \ \tilde\mu=\infty, \ \ \ \ \ 
\tilde\zeta=0
\ee
is the high temperature fixed point. Here the infrared 
properties 
are
determined by the charged plasma effects.

3. The point $T_{\rm GG}$
 \be
t=1,\ \ \ \ \ \tilde\mu=\tilde\zeta, \ \ \ \ 
\tilde\mu=\infty \ee
is the infrared unstable fixed point. This is precisely the 
critical
point
 that corresponds to the deconfining phase transition. At 
this point
 both the monopole and the charge plasma induced terms are 
equally
important.

Note that the fixed line $t>2, \ \ \tilde\zeta=0$ of 
massless 
theories
which corresponds to
the ``confined monopole plasma'' is not present, since for 
$t>2$ the
charged plasma induced mass term is strongly relevant. The 
same is
true for the would be fixed line $t<1/2, \ \ \tilde\mu=0$, 
which is
present in the absence of the monopoles and describes the 
low
temperature massless phase of noncompact electrodynamics.

It is instructive to see how the renormalization group 
equations formalize our
qualitative arguments of the previous section. In 
particular, 
they
make clear the role of the point $t=1/2$ which, in the 
absence of
monopoles, would be the point where the charged induced term
becomes relevant. If one starts the evolution from the 
initial
condition $t=1/2$, $\tilde\mu\gg\tilde\zeta$, the running
temperature will increase, and $\tilde\mu$ will grow into 
the
infrared, while $\tilde\zeta$ will grow for a while until 
the
running temperature reaches two. From this point on, 
$\tilde\zeta$
will decrease and the system will approach the high 
temperature
fixed point. This corresponds to the situation where at the
would be critical point $t=1/2$, the density of charged 
plasma
is so high that the mean distance between the particles is 
smaller
than the width of the confining string.

On the other hand, if the initial condition is
$\tilde\mu<\tilde\zeta$, the temperature starts decreasing, 
making
the coupling $\tilde\mu$ immediately irrelevant. Thus, 
$\tilde\mu$
monotonically decreases to zero and does not significantly 
affect
the flow of the other couplings, which steadily flow to the 
zero
temperature fixed point. This holds in the situation where 
at
$t=1/2$ the density of charged plasma is low, which is 
indeed true
in our model.

Interestingly enough, at $t=1$ the initial conditions in our 
model
in the case of a very heavy Higgs are such that the 
renormalization group flow 
starts
almost exactly in the region of attraction of the fixed 
point
$T_{\rm GG}$. At this value of temperature the monopole and 
charge
fugacities are equal. The only difference between $\tilde 
\mu$ and
$\tilde\zeta$ is then in the prefactors, which are possibly
different combinations of $M_W$ and $g^2$. If $\tilde\mu$ 
and
$\tilde\zeta$ were exactly equal, the only thing that 
happens
along the flow is that their values grow, but $t$ does not 
change.
Due to the small difference their values are equal at a
temperature which slightly differs from $1$, 
\be {M_W\over
T}={4\pi M_W\over g^2}+O(1), \ \ \ \ \ \ T={g^2\over
4\pi}\left[1+O\left({g^2\over M_W}\right)\right]. 
\ee
 Thus, the initial 
temperature for
which the system is in the region of attraction of $T_{\rm 
GG}$  is
slightly different. We conclude that the critical 
temperature of
the Georgi-Glashow model is indeed given by $T_{\rm 
GG}=\frac{g^2}{4\pi}$, up to corrections of 
order
$g^2/M_W$.

\subsection{The universality class of the phase transition}

Since the deconfining phase transition is due to the 
restoration
of $Z_2$ symmetry, we expect that it is in the 
two-dimensional Ising
universality class. This can be shown exactly by studying 
the
theory (\ref{sinevortex}) at the fixed point $t=1$. The
following discussion follows closely Ref. 
\citebk{Nersesyan}. [See 
also
Chapter 21 in the book \citebk{Tsvelik}].

The theory described by Eq.~(\ref{sinevortex}) can be 
fermionized
by using the standard bosonization/fermionization 
techniques.
Since at $t=1$ both cosine terms have dimension $1$, the 
resulting
fermionic theory is a theory of free massive fermions. In 
our
notations, the Dirac fermionic field is defined as \be
\psi_{R}=a^{-1/2}i\exp
\left[I\left(\chi+ {\tilde\chi\over 2}\right)\right], \hskip 
1 cm
\psi_{L}=a^{-1/2}\exp
\left[- I\left(\chi- {\tilde\chi\over 2}\right)\right] .
 \ee 
The
kinetic term in Eq.~(\ref{sinevortex}) then becomes the 
kinetic
term of the field $\psi$, while the cosine terms become
\begin{eqnarray}
&&a^{-1}\cos 2\chi=i[\psi_R^\dagger\psi_L-{\rm 
H.c.}],\nonumber\\[2mm]
&&a^{-1}\cos \tilde\chi=i[\psi_R^\dagger\psi^\dagger_L-{\rm 
H.c.}].
\end{eqnarray}
Thus, the mass term in the fermionized Lagrangian is 
diagonalized
by introducing the real Majorana fermions
\be
\rho={\psi+\psi^\dagger\over \sqrt 2}, \ \ \
\sigma={\psi-\psi^\dagger\over i \sqrt 2}.
\ee
The mass of the fermion $\rho$ is $\mu a+\zeta a$, while
the mass of $\sigma$ is $\mu a-\zeta a$.
For $\mu=\zeta$ the model contains one massive and
one massless fermion. At the fixed point 
$\tilde\mu\rightarrow\infty$
the massive fermion decouples. Thus, at the point $T_{\rm 
GG}$ 
the theory
is that of one massless Majorana fermion. It is well know 
that this
theory precisely describes the critical point of the 2D 
Ising model.

Thus, we indeed see that the phase transition is in the 
universality
class of the 2D Ising model.

\section{Vortices and monopoles at high temperature}

It is interesting to see how the behavior of magnetic 
vortices and
the monopoles is affected by the deconfining phase 
transition. We
will do this with the help of the effective action written 
in
terms of the zeroth component of the vector potential $A_0$. 
The
relation between our analysis, which was performed for the 
phase
of the vortex operator $\chi$ and this standard procedure is 
not
difficult to see. The free energy of the charged $W^\pm$ 
bosons in
the Lagrangian (\ref{sinevortex}) is given by the term
$\cos\tilde\chi$. On the other hand, this same free energy 
in 
the
standard calculation is represented by the insertion of the
Polyakov line with charge two. Thus, we identify the 
Polyakov 
line
with the exponential of the dual field 
\be P=\exp\left( {i\over
2}\tilde\chi\right)\,, \ee 
and the dual field $\tilde\chi$ with the
Abelian vector potential \be \tilde\chi=2g\beta A_0. 
\label{iden}
\ee 
Up to the monopole induced term, the Lagrangian
(\ref{sinevortex})  is equivalent to the sine-Gordon 
theory of 
the
field $\tilde\chi$ in Eq.~(\ref{dualsG}). With the 
identification
in Eq.~(\ref{iden}), this is 
\be {2\over 
T}(\partial_iA_0)^2+ \mu
\cos \left({2g\over T}A_0\right). 
\ee
 The monopole induced term also 
can be
written in terms of the vector potential. Its form in terms 
of the
Polyakov line $P$ is precisely the same as the form of the 
charge
plasma induced term in terms of the vortex field $V$. Recall 
that
the origin of the plasma induced term $\cos\chi$ in Eq.
(\ref{sinevortex}) is the dimensionally reduced ``Skyrme 
term'' of 
Eq.~(\ref{skyrme}). Thus, just like we derived the 
Sine-Gordon
Lagrangian (\ref{sinevortex}) starting from the 
effective
Lagrangian for the vortex field, we can follow the same 
steps
backwards, but this time expressing everything in terms of 
$P$.
Due to this duality between $V$ and $P$ we conclude that the
monopole induced term is the dimensionally reduced Skyrme 
term for
the Polyakov line. The full effective Lagrangian for the 
vector
potential is then 
\be
{\cal L}={2\over T}(\partial_iA_0)^2+\mu \cos
\left({2g\over T}A_0\right)+ {a^2\over 4\pi^2}\ln 
\left(\tilde\zeta
\right)(\epsilon_{ij}
\partial_iP\partial_jP^*)^2.
\label{a0}
\ee
The monopole in this Lagrangian is represented by a vortex 
of the
field ${g\over T}A_0$ with unit vorticity.
The coefficient of the Skyrme term in Eq.~(\ref{a0}) is such 
that 
the
action of such a unit vortex is equal to the action of the 
``core'' 
of
the monopole.

The cosine term in this expression is the potential for 
$A_0$
induced by the nonvanishing density of the charged particles 
in
the thermal ensemble. Naturally, it contains the Debye
``electric'' mass term for $A_0$ and also higher 
interactions.
Note that, as opposed to strongly interacting theories, 
where
similar effective Lagrangians have been derived only in the
derivative expansion, the Lagrangian (\ref{a0}) is valid 
on
all distance scales longer than $1/T$. Thus, in principle, 
it can be
used to calculate correlation functions in a large momentum 
range.
However, if one is interested in the long distance behavior 
of the
correlators, at temperatures above $T_{\rm GG}$, the higher 
derivative
Skyrme term can be neglected. This is in accordance with our
analysis of the previous section, which showed that the 
monopole
induced term is irrelevant above $T_{\rm GG}$.

Consider now the correlation function of two vortex 
operators.
Recall that in the confining phase the vortex operator has a
nonvanishing expectation value  since the magnetic $Z_2$
symmetry is spontaneously broken. In the deconfined phase we 
expect
the correlation function of two vortex fields to decay
exponentially at large distance.

Consistent with the magnetic symmetry breaking, at zero
temperature we have 
\be \langle V(x)V^*(y)\rangle={g^2\over
8\pi^2}\exp\left[-{1\over 2}\langle\chi(x)\chi(y)\rangle\right]= 
{g^2\over
8\pi^2}\exp\left[-{16 \pi^2\over g^2}Y_3(x-y)\right] \,,\label{t0} 
\ee 
where
$Y_3(x-y)$ is the 3d Yukawa potential with the mass $M$. At
temperatures below the phase transition ($T<T_{\rm GG}$) but 
high
enough so that the dimensional reduction is valid the 
infrared
asymptotics (at distances $|x-y|>1/T$) is instead 
\be
\langle V(x)V^*(y)\rangle={g^2\over 8\pi^2}\exp\left[-{16 
\pi^2T\over
g^2}Y_2(x-y)\right] \label{lowt} 
\ee 
with $Y_2(x-y)$  the 
two-dimensional Yukawa
potential with the mass which includes the effects of 
integrating
out the nonzero Matsubara modes, as calculated in 
Ref.~\citebk{zarembo}.
This expression follows from Eq. (\ref{sinevortex}) 
neglecting the
$\cos\tilde\chi$ term, which is indeed negligible in the 
infrared.

At high temperatures $T>T_{\rm GG}$ it is convenient to use 
Eq.~(\ref{a0}) with the omission of the third term. The 
calculation of
the correlator of the vortex operators proceeds along the 
lines of
Refs. \citebk{kaks}, \citebk{Kovner1}. The insertion of 
$V(x)$ and 
$V^*(y)$ 
creates
the $Z_2$ domain wall stretching between the points $x$ and 
$y$.
In terms of the sine-Gordon theory, Eq. (\ref{a0}) this 
domain
wall is just the kink, and thus the domain wall tension is 
equal
to the soliton mass $M_s$.
 The correlator is then \be
\langle V(x)V^*(y)\rangle=\exp\{-M_s|x-y|\} \label{ms} \ee 
with the 
soliton
mass \be M_s= a^{-1}{2\Gamma(p/2)\over 
\sqrt{\pi}\Gamma({{p+1\over
2}})}\Bigg[{ \pi \Gamma({{1\over p+1}})\over \Gamma({{p\over
p+1}})}\, 2\tilde\mu \Bigg]^{{p+1}\over 2}, \ee where \be p= 
{
g^2\over 8\pi T - g^2}. \ee

Thus, we find that the correlation function of the vortex 
operators
decreases exponentially in the high temperature phase as it 
should
in the phase with restored symmetry. Note that, were we to 
neglect
the Debye mass term, we would have found that below 
$T_{\rm BKT}$ the
correlator tends to a constant at infinity, while at 
$T>T_{\rm BKT}$
it decays at large distances, but only as a power, since the 
mass
of the soliton vanishes.

What about the monopoles? The discussion in the beginning of 
this
section would suggest that they are ``confined" 
logarithmically.
However, it is easy to see from the effective action 
Eq.~(\ref{a0})
that the monopoles are bound by a (screened) {\it linear}
potential. Consider a configuration with unit winding of the 
field
$P$. Due to the potential term $P^2$, the minimal action
configuration cannot be a rotationally symmetric hedgehog. 
Such a
configuration would ``cost" action proportional to the 
volume,
since the field $P$ would be away from its vacuum value 
everywhere
in space. The best one can do is to have a quasi one 
dimensional
strip in which the winding is concentrated, while everywhere 
else
in space $P$ would be equal to $1$ (or $-1$). This 
configuration
is schematically depicted in Fig.~4.

\begin{figure}
\hskip 4cm
\psfig{figure=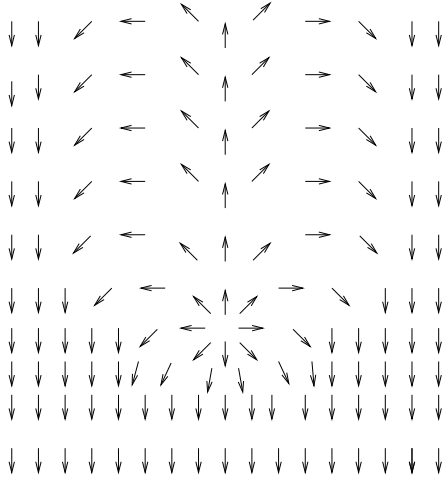,height=2.5in}
\caption{The  string-like
configuration of the Polyakov line $P$ that accompanies the
monopole-instanton in the high temperature phase.}
\label{fig:Fig.4}
\end{figure}

The width of the ``dual confining string" must clearly be of 
order
of $M^{-1}_D$, while the action per unit length, $s\propto 
{T\over
g^2}M_D$. Thus, the action of a single monopole diverges 
linearly
with the size of the system. Obviously the 
monopole-anti-monopole
pair separated by a distance $L$ have action $Ls$. When the
distance is large enough, another pair can be produced from 
the
thermal ensemble, screening the linear potential. The 
critical
distance at which this happens is determined by \be
L_cs=2\log\zeta={8\pi M_W\over g^2}, \ee or \be L_c\propto
{M_W\over T}M^{-1}_D. \ee Thus, as long as the temperature 
is much
lower than $M_W$, the length of the ``dual string" is much 
greater
than its thickness. The ``potential" between the monopoles 
is
therefore linear, but screened at large distances, much the 
same
as the confining potential in gauge theories with heavy
fundamental charges. Thus, the behavior of the monopoles is 
in many
senses ``dual'' to that of electric charges\footnote{Of 
course one
always has to keep in mind that physically monopoles and 
charges
are very different objects in this model: the charges are
particles, while the monopoles are instantons.}.

Even though the monopoles are ``confined", their effects do 
not
disappear in the hot phase. They, for example, are the main
contributors to the correlation functions of the vortex 
operators,
of the type $V(x)V(y)$. This correlator would be vanishing 
if the
magnetic symmetry were U$(1)$ rather than $Z_2$. It was 
shown,
however, in Ref. \citebk{inst}, that this correlation 
function does 
not
vanish due to the monopole contributions. In fact, the 
``diagonal"
correlation functions are those of the parity even and 
parity odd
operators $V^+=V+V^*$ and $V^-=V-V^*$:
 \bea
\langle V^+(x)V^+(y)\rangle&=&(aM_D)^{\pi T\over g^2}
\exp\{-|x-y|\tilde\sigma_-\}\nonumber \\[2mm]
\langle V^-(x)V^-(y)\rangle&=&(aM_D)^{\pi T\over
g^2}\exp\{-|x-y|\tilde\sigma_+\} \eea with \be 
\tilde\sigma_\pm=
M_s\pm\zeta(aM_D)^{{2\pi T\over g^2}-1}{1\over \pi a} .\ee 
Thus,
the correlation lengths in the scalar and pseudoscalar 
channels
are not equal due to the monopole effects. At high 
temperatures
this difference becomes small, since $\zeta\ll aM_D\ll 1$, 
but 
it
nevertheless remains strictly finite at any finite 
temperature.

\section{Hot confining strings:
Deconfinement versus Hagedorn transitions}

In this section we discuss the deconfining phase transition 
from
the perspective of the confining string.

What happens to strings at very high temperatures is one of 
the
big
 questions of string theory. In the early days of  dual 
resonance 
models and
 hadronic bootstrap, the exponential growth in the 
one-particle 
density of
states \beq \label{hag} \rho(m) \sim  m^a \,\,e^{bm} \eeq 
led to
the famous Hagedorn transition.\cite{HAGEDORN} This type of
spectrum first arose in the context of statistical bootstrap
models\,\cite{HAGEDORN,FVHW,FC} and, for
 hadrons, such behavior indicates that they are composed of 
more
fundamental constituents.\cite{cab} In fundamental string
theories one finds  the same kind of spectrum (see, for 
example,
books \citebk{BOOKS}), and a search for hints to the 
existence of 
``string
constituents² is of great interest. What lies beyond the 
Hagedorn
temperature? Is this temperature limiting or is there a
high-temperature phase which reveals the fundamental degrees 
of
freedom?  And is it true that for all types of string 
theories
there is the same universal physics  or is it that different
classes may have totally different high-temperature 
behavior? A
lot of effort has been invested into study of the Hagedorn
transition in critical (super)strings. There is an enormous 
volume
of literature on this subject, some references (but by no 
means
all) can  be found, for example, in Ref. \citebk{ABKR}.
 For weakly coupled critical (super)strings  the Hagedorn 
transition  
can be
described as a Berezinsky-Kosterlitz-Thouless\,\cite{BKT}
transition on a 
world sheet.\cite{KOGAN87,SATHIAPALAN87,ATICK,AK} 
It is due to the ``condensation" of the world sheet 
vortices. It
has been also suggested that the transition in some cases  
may
actually be first order.
 Above the Hagedorn temperature, the vortices
populate the world  sheet and  we have a new phase. From the 
target
space point of view we are talking about tachyonic 
instabilities
for non zero winding modes in the imaginary time direction
(thermal winding modes).

In the more general case of interacting strings, one does 
not know
with certainty what the fundamental degrees of freedom are 
and
thus what their role is at high temperature. Recently it was
suggested, in the framework of the Matrix Model description 
of the
Hagedorn transition, that the fundamental string decays into 
D0
Branes.\cite{SATH}  Thus, it could be that D0 branes are the
fundamental degrees of freedom in the hot phase.

In short, there are many interesting open string theory 
questions
pertaining to the hot string. Since our simple 2+1 
dimensional
model is well understood, it is worth seeing what it has to 
say on
these issues.

\subsection{The monopole binding as the Hagedorn transition}

As we saw in the previous sections, the dynamics of the
deconfining phase transition is quite nontrivial, involving
interplay between the charged degrees of freedom and the
monopoles. To disentangle these effects in  string language, 
we
start our discussion by
 completely disregarding the charged
particles. From the string point of view, this means that we
neglect possible contributions of the heavy D0 branes, and 
are
thus entirely within the theory of closed strings. Naively 
one
expects in such a theory the existence of a Hagedorn 
temperature,
beyond which the string cannot exist. In an almost free 
string
this temperature is of order of the string scale. One can
visualize this phenomenon in simple terms. Consider a closed
string of a given fixed length $L$. Let us calculate the 
free
energy of such a string. The energy of the string is
\begin{equation}
E=\sigma L
\end{equation}
The number of states for a closed string of length $L$ 
scales
exponentially with $L$
\begin{equation}
N(L)=\exp\{\alpha L\}.
\end{equation}
The dimensional constant $\alpha$ is determined by the 
physical
thickness of the string. Imagine that the string can take 
only
positions allowed on a lattice with the lattice spacing $a$. 
Then
clearly the number of possible states is $z^aL$, where $z$ 
is a
number of order unity, equal to the number of nearest 
neighbors on
the lattice.\footnote{We disregard in this argument the fact 
that
the string has to close in on itself. This extra condition 
would
lead to a pre-factor with power dependence on $L$. Such a 
pre-factor
is not essential for our argument, and we therefore do not 
worry
about it.} In this simple situation $\alpha=a\ln z$. The 
only
natural lattice spacing for such a discretization is the 
thickness
of the string. For an almost free string the thickness is
naturally the same as the scale associated with the string
tension. Thus, the entropy is
\begin{equation}
S=x\sqrt\sigma L
\end{equation}
with $x$ a number of order unity. The free energy is then
\begin{equation}
F[L]=\sigma L-xT\sqrt\sigma L
\end{equation}
At the temperature
\begin{equation}
T_H={1\over x}\sqrt\sigma
\end{equation}
the free energy becomes negative, which means that strings 
of
arbitrary length appear in the thermal ensemble in a 
completely
unsuppressed way. The thermal vacuum becomes a ``soup'' of
arbitrarily long strings. Thus, effectively, the 
``temperature
dependent'' string tension vanishes and it is not possible 
to talk
about strings any more in the hot phase.
 For more details about this ``random walk'' description of 
hot
strings, see Ref. \citebk{hotsoup} and references therein.

The situation is very similar to the BKT phase transition, 
where the 
free
energy of a vortex becomes negative at the critical 
temperature, and 
the
vortices populate the vacuum in the hot phase.

In the string partition function language this is just a
restatement of the well known fact that the partition 
function
diverges in a sector with the topology of a torus which 
winds
around the compact Euclidean time direction. Fixing the unit
winding in the Euclidean time physically corresponds to the
calculation of the free energy in the sector with one closed
string. The integration over all possible lengths of the 
string is
the cause of the divergence of the partition function at 
high
temperature.

The same physical effect exists for the confining Georgi-
Glashow string. 
There
is one difference, which however turns out to be crucial for 
the
nature of the phase transition. This is that the thickness 
of the
Georgi-Glashow string is not given by the string tension. 
Rather it is 
equal
to the inverse photon mass, and thus
\begin{equation}
\alpha\propto M_\gamma\propto {\sigma\over g^2}\ \ \ .
\end{equation}
We thus have for the confining Georgi-Glashow string
\begin{equation}
T_H={\sigma\over \alpha}\propto g^2.
\end{equation}
This is indeed the correct magnitude of the critical 
temperature
as discussed in previous sections.

The noteworthy feature of this formula is that the Hagedorn
temperature of the confining string is much higher than the 
string
scale. This is easy to understand because the entropy of the 
thick
string is much smaller than that of the free string due to 
the
fact that high momentum modes of the confining string do not
contribute to the entropy at all, as discussed earlier. 
Thus, one
needs to heat the string to a much higher temperature for 
entropy
effects to become important.

Since we have completely neglected the possibility of the
appearance of D0 branes (charged particles), the 
transition we
have been discussing is the string representation of the 
monopole
binding transition in the Georgi-Glashow model. At the point 
where
the monopoles bind, the photon of the Georgi-Glashow model 
becomes 
massless
and thus the string tension disappears and the thickness of 
the
string diverges. In the Hagedorn picture this is just the 
dual
statement that the ensemble is dominated by infinitely long
strings. The BKT nature of both transitions underlines this 
point.

\subsection{Vortices on the world  sheet  ---  open strings 
and charged 
particles}

Sometimes the Hagedorn transition is discussed in terms of 
the
vortices on the world  sheet. What is the physical nature of 
these
objects?

Consider a string world  sheet with the topology of a sphere 
with a
vortex-anti-vortex pair, Fig.~5. When going around the 
vortex
location on the world  sheet, the compact coordinate $x_0$ 
varies
from $0$ to $\beta$. Since this is true for any contour of
arbitrarily small radius, which encircles the vortex, this 
means
that physically the location of the vortex in fact 
corresponds in
the target space not to a point but rather to a line which 
winds
around the compact direction. The vortex-anti-vortex pair on 
the
world  sheet thus represents an {\it open} string which 
winds
around the compact direction. Figure 5 illustrates how the 
open
string world  sheet, equivalent to a cylinder is transformed 
by a
conformal transformation into a sphere with two singular 
points  --- 
the vortex-anti-vortex pair.

\begin{figure}
\hskip 4cm
\psfig{figure=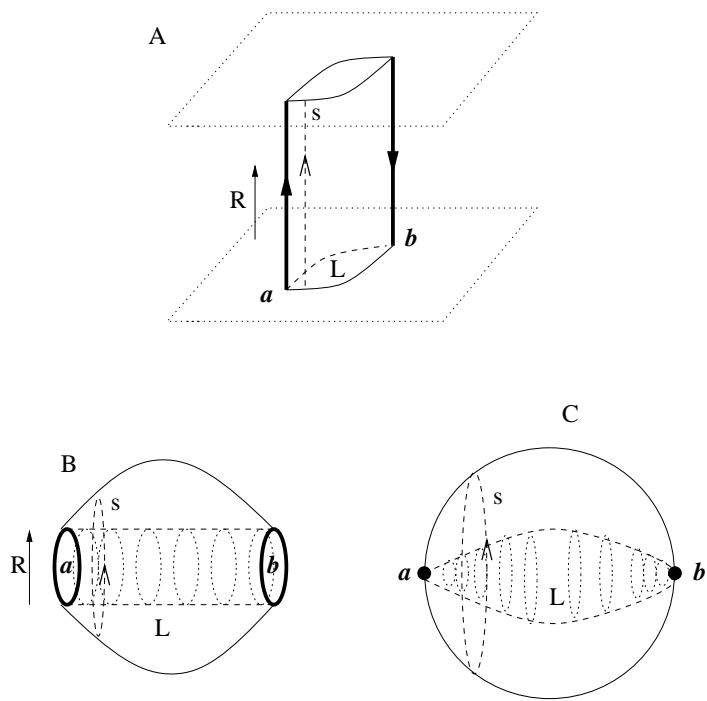,height=4.5in}
\caption{ Representation of a vortex and an anti-vortex
pair with a string stretched between them and the
corresponding string world sheets.}
\label{fig:Fig.5}
\end{figure}

If the string theory in question does have an open string 
sector,
then the configurations with arbitrary number of
vortex-anti-vortex pairs contribute to the finite 
temperature
partition function. The Hagedorn phase transition can then 
be
discussed in this sector rather than in the closed string 
sector.
Not surprisingly, the discussion is exactly the same as in 
the
previous subsection. The fact that we now have an open 
rather than
closed string does not change the entropy versus energy 
argument.
One finds that at $T<T_H$ the vortices on the world  sheet 
are
bound in pairs. The corresponding target space picture is 
that the
``ends" of the open strings are bound by a linear potential 
and
therefore long open strings do not contribute to the thermal
ensemble. In exactly the same way, long closed strings are 
also
absent from the ensemble. At $T>T_H$ the vortices unbind and
appear in the ensemble as a Coulomb gas. Thus, a typical
configuration contains lots of open strings (as well as lots 
of
arbitrarily long closed strings) since the energy of such 
strings
is overwhelmed by the entropy.

Note, however, that the existence of the transition in this
context is entirely independent of the presence or absence 
of
vortices. As we saw above, the transition can be understood 
purely
on the level of the closed string. It is driven by the 
string
fluctuations. Thus, even if the theory does not have an open 
string
sector, the transition is still there. For example there are 
no
open strings in the  Georgi-Glashow model. Nevertheless, if 
we
neglect the effects of $W^\pm$ the Hagedorn transition is 
still
there and it coincides with the ``monopole binding" 
transition.

The actual phase transition in the Georgi-Glashow model is 
however
not driven by monopole binding. Nevertheless, in the string
language
 it is also due to proliferation of vortices on the
world sheet. These are, however, vortices of a somewhat 
different
type. The same conformal transformation that turned the open
string boundary into a point can be used to turn  the 
world  line
of a  heavy particle into a point  ---  let's call it $0$ 
brane. As
discussed above, this is a (fundamentally) charged particle 
which
couples to the Georgi-Glashow confining string. Thus, a 
string 
world  sheet
representation of a pair of particles with opposite charge 
is also
a vortex-anti-vortex pair on a sphere.

There are some important differences between these vortices 
and
the ones that represent the ends of the open string. First, 
for an
open string the noncompact ``spatial" coordinates satisfy 
Neumann
boundary conditions. Thus, even as $x^0$ winds, the other
coordinates $x^i$ can take arbitrary values close to the 
vortex.
One can see this from the action
\begin{equation}
S = \sigma \int d^2 \xi \partial_a x^0 \partial_a x^0 +
\sigma \int d^2 \xi \partial_a x^i \partial_a x^i
\end{equation}
where the dynamics of $x^i$ is absolutely unrelated to the
dynamics of $x^{0}$. Thus, even if a vortex in $x^0$ sector  
can
be considered as a boundary in a target space,  there are 
free
boundary conditions for the components $x^i$. In other 
words,  we do
not have any boundary action induced by vortices in a theory 
of closed strings.

 For a very heavy $0$ brane the boundary conditions
are of the Dirichlet type. Thus, $x_i$ are constant close 
enough
to the vortex location. Another difference is that, since 
the $0$
brane is an independent degree of freedom, in principle its 
mass
is a free parameter. Thus, the fugacity of the $0$ brane 
vortex is
an independent parameter, and physics may in principle 
depend on
it. Of course, in the case when we have a finite mass for 
$0$
branes there is a nontrivial boundary action describing the
massive particle. In that case one has  non-conformal 
boundary
conditions compatible with a finite mass of a $0$ 
brane.\cite{KW}

The Georgi-Glashow confining string has neither open string
sectors nor $0$ branes which are the sources of a single 
string.
The dynamical objects which couple to the string are the 
charged
$W^\pm$ particles, which have an adjoint charge and 
therefore are
sources of a pair of strings.

Each dynamical $0$ brane in the Georgi Glashow model has two
strings emanating from it. Thus, a pair of
 branes
propagating in compact imaginary time is conformally 
equivalent to a
 pair
of ``double vortices'' as in Fig.~6.

\begin{figure}
\hskip 4cm
\psfig{figure=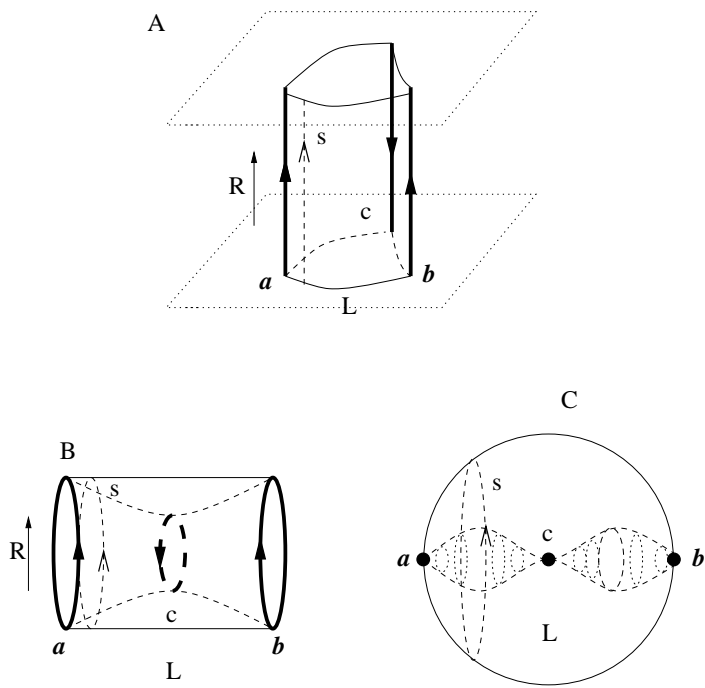,height=4.5in}
\caption{See the text}
\label{fig:Fig.6}
\end{figure}

The singular points are still vortices as before, since 
going
around such a point one travels once around the compact time
direction. But now two string world  sheets are permanently 
glued
together at the location of a vortex. The fact that two
world  sheets are glued at the location of the vortex is the
manifestation of the $Z_2$ magnetic symmetry of the Georgi-
Glashow 
model. The
region of space between the two world  sheets is separated 
from the
region of space on the outside, reflecting the fact that it
constitutes a domain of a different vacuum.

At low temperatures such configurations in the thermal 
ensemble
are rare. But at the critical temperature their density 
becomes so
large that the ensemble is dominated by multi-vortex
configurations. In the string picture these are 
configurations
with multiple points of ``gluing''. Importantly, the $0$ 
branes
are not just pairwise connected by two strings, but, rather, 
form a
network where all of them are connected to each other. 
Clearly in
terms of entropy such configurations are much more 
favorable, and
there is also no loss of energy when the distance between 
the $0$
branes is of the order of the string thickness.

This string-based description of the transition has to be 
taken
with a grain of salt. The relevant physics takes place on 
short
distance scales  ---  of the order of the thickness of the 
string. On
these scales, as we explained above, the string modes are
practically absent. Thus, even though we have showed the 
string
segments on Fig.~6, these segments are so short that there 
is no
string tension associated with them. Thus, the mechanism of 
the
transition is essentially field-theoretical rather than
string-theoretical.

To understand this point better consider in a little more 
detail
the thermal ensemble of $W^\pm$. The crucial point is that 
at
distance scales $d\ll M_\gamma^{-1}$ the interaction between 
them
is Coulomb rather than linear. The gas of charges with 
Coulomb
interaction has itself a transition into a plasma phase. 
This
transition has nothing to do with the long range linear
interaction and occurs at the temperature $T_{NC}$ which is 
four
times smaller than the Hagedorn temperature.\cite{dunne} At 
this
temperature the $W^\pm$ become ``free" in the sense that 
they cease
to care about the Coulomb part of the potential. The crucial
question however is: how large is the density of $W^\pm$ at 
this
point? If the density at $T_{NC}$ were large and the average
distance between $W$'s were smaller than $M_\gamma^{-1}$, 
the
transition would actually occur at $T_{NC}$, since the  long 
range
linear part of the potential would be entirely irrelevant. 
As it
happens, in the Georgi-Glashow model this is not the case, 
and the 
density of
$W$ is small. Thus, at $T_{NC}$ there is a certain 
rearrangement of
the thermal ensemble on short distance scales, but at long
distances nothing happens  ---  the string still confines. 
However, at
$T_C=T_H/2$ the density of $W$ reaches the critical value 
and the
transition occurs. Note that at this temperature the large 
length
fluctuations of the string are still suppressed  ---  we are 
far below
the Hagedorn temperature. The string is destroyed not due to
``stringy" physics of the Hagedorn transition, but due to 
the short
distance field-theoretical effects, viz. the fact that the
fugacity of $W$ is relatively large and that the interaction 
at
short distances is Coulomb and not linear.

\section{The SU$(N)$ model}
So far in this review we have discussed the SU$(2)$ gauge 
theory.
Most of the discussion is easily generalized to the weakly
interacting theory with SU$(N)$ gauge group,
 \bea {\cal{L}}=
-{1\over 2} \mbox{tr}F_{\mu \nu}F^{\mu \nu} +
 \mbox{tr}D_\mu \Phi D^\mu \Phi  -  V(\Phi)\,,
\label{modeln} \eea
 where
\bea
A_\mu & = & A^a_\mu T^a, ~~~~~
F_{\mu\nu}  =
\dd_\mu A_\nu  -\dd_\nu A_\mu +  g[A_\mu, A_\nu], \nonumber 
\\[3mm]
\Phi & =  & \Phi^a T^a, ~~~~~ D_\mu \Phi  =  \dd_\mu \Phi + 
g[A_\mu, \Phi]. 
\eea 
Here $T^a$ are traceless Hermitian
generators of the SU$(N)$ algebra normalized such that
$\mbox{tr}(T^a T^b)={1\over 2}\delta^{ab}$.

Depending on the form of the Higgs potential, there can be
different patterns of gauge symmetry breaking. Since most of 
the
details of the potential are unimportant for our purposes, 
we will
not specify it except for restricting it to the region of 
the
parameter space where classically the gauge symmetry is 
broken to
the maximal torus 
\be {\rm SU}(N) \rightarrow {\rm U}(1)^{N-1}.
\label{symmetry} \ee We also restrict ourselves to the 
weakly
coupled regime, which means that the ratios $M_W/g^2$ are 
large
for {\it all} $N^2-N$ massive $W$ bosons.

 To characterize the perturbative
spectrum of the theory it is convenient to use the 
Cartan-Weyl
basis $(H^i, E^{\vec{\alpha}})$, where $H^i$ generate the 
Cartan
subalgebra which is of dimension of the rank  $r= N-1$ of 
SU($N$),
\bea [H^i,H^j]=0\,,  \hskip 1 cm i,j \in [1,2,.. N-1]\,,
 \eea 
and
$E^{\vec{\alpha}}$ are the $N(N-1)$ ladder operators which 
satisfy
\bea
&[H^i, E^{\vec{\alpha}}] & =  \alpha^i E^{\vec{\alpha}}, \\[2mm]
& [E^{\vec{\alpha}},E^{\vec{\beta}}] & =
N_{\vec{\alpha},\vec{\beta}}\,\, 
E^{\vec{\alpha}+\vec{\beta}}
\hskip 0.5 cm \mbox{if}\,\,\, \vec{\alpha}+\vec{\beta} 
\,\,\,\mbox{is
a root}  \\[2mm]
& ~~ & =  2\vec{\alpha} \cdot \vec{H} \hskip 0.5 cm 
\mbox{if}
\,\,\,\,\vec{\alpha} = - \vec{\beta}. 
\eea 
The $(N-1)$-dimensional
root vectors  $\vec{\alpha} = (\alpha^1,\alpha^2,...
\alpha^{N-1})$ form the dual Cartan subalgebra. There are
obviously $N(N-1)$ such vectors corresponding to
$\mbox{dim(SU}(N))- \mbox{rank(SU}(N))$ but only $N-1$ of 
them are
linearly independent. The non-vanishing inner products in 
the
Cartan-Weyl basis are \be \mbox{tr}(H^i,H^j) = {1\over 
2}\delta^{i
j}, \hskip 1 cm \mbox{tr}(E^{\vec{\alpha}},E^{\vec{\beta}}) 
=
{1\over 2}\delta^{\vec{\alpha},-\vec{\beta}}. \ee

At the classical level $N-1$ gauge group generators are 
unbroken,
which we choose to correspond to $\{H^i\}$. Therefore, 
classically
there are $N-1$ massless photons and $N(N-1)$ charged 
massive
$W$ bosons.

Our Weyl basis is chosen in such a way that the Higgs VEV is
diagonal. Since the matrix $\Phi$ is traceless, there are 
$N-1$
independent eigenvalues. In terms of the $N-1$ dimensional 
vector
$\vec{h}= (h_1, h_2, h_3,..h_{N-1})$
we 
have\,\footnote{ For 
concreteness we
order these numbers $h_1 \,\,> \,\, h_2 > ... h_{N-2} 
\,\,>\,\,
h_{N-1}$, which also breaks the discrete Weyl group.}  
\be
\langle\Phi\rangle =  \vec{h} \cdot \vec{H} , \ \ \ \ \ \ 
A_\mu = 
\vec{A}_\mu
\cdot \vec{H} + \sum_{\vec{\vec{\alpha}}} 
A^{\vec{\alpha}}_\mu
E^{\vec{\alpha}} \,.
\ee 
For concreteness let us choose the 
following
basis for the Cartan subalgebra: \bea
&&H_1= {1\over 2}\mbox{diag}(1,-1, 0, ...0), \hskip 0.5 cm 
H_2= 
{1\over 
2\sqrt{3}}\mbox{diag}(1,1, -2, 0 ...0) \nonumber \\
&& \,\,\,\, ... \ \ \ H_{N-1}= {1\over\sqrt{2N(N-1)}}
\mbox{diag}(1,1, 1, ...1, - (N-1)). \eea As long as $\vec{h} 
\cdot
\vec{\alpha} \neq 0$ for all roots, the gauge symmetry is
maximally broken. The masses of the $W$ bosons can be read 
off from
the second term in the Lagrangian 
\bea && g^2
\mbox{tr}[A_\mu,\Phi]^2 = {g^2\over 
2}\sum_{\vec{\alpha},i,j}
A^{\vec{\alpha}}_{\mu}A^{-\vec{\alpha}}_{\mu}h_i h_j 
\alpha^i 
\alpha^j \\
&&\Longrightarrow  M_{\vec{\alpha}} = g |\vec{h} \cdot
\vec{\alpha}|. \eea

The $W$ bosons corresponding to the $N-1$ simple roots 
$\vec\beta_i,
\ \ \ i=1,...,N-1$ (an arbitrarily chosen set of linearly
independent roots) can be thought of as fundamental, in the 
sense
that the quantum numbers and the masses of all other $W$ 
bosons are
obtained as linear combinations of those of the fundamental
$W$ bosons. These charges and masses are \be 
\vec{Q}_{\vec{\beta}}=
g\vec{\beta}, \hskip 1 cm M_{\vec{\beta}} = g \vec{h} \cdot
\vec{\beta}\,. \label{cham} \ee As an example, consider the 
case of
SU$(3)$ broken down to U$(1)\times U(1)$. There are 6 
massive
$W$ bosons. The simple roots can be taken as \be 
\vec{\beta}_1 =
({1\over 2}, {\sqrt{3}\over 2}), \hskip 1cm \vec{\beta}_2 =
(-{1\over 2}, {\sqrt{3}\over 2}) \,.\label{simpleroots} \ee 
The
remaining non-simple positive root is \be \vec{\alpha}_3 =
\vec{\beta}_1 - \vec{\beta}_2 = (1,0). \ee The other three 
roots
are $-\vec\beta_i$, $-\vec\alpha_3$. The masses of 
corresponding
$W$ bosons are \be M_{W_1}= {g\over 2}(h_1 +\sqrt{3} h_2 ), 
\hskip
0.5 cm M_{W_2}= {g\over 2}(h_1 -\sqrt{3} h_2 ), \hskip 0.5 
cm
M_{W_2}= gh_1 \label{wmass} \ee for $h_1 > \sqrt{3} h_2$.

\subsection{The monopole-instantons and the Polyakov 
effective 
Lagrangian}
Non-perturbatively, the most important contributions in the 
theory
are due to the monopole-instantons. Those are classical, 
stable,
finite action solutions of the Euclidean equations of motion
arising due to the nontrivial nature of the second homotopy 
group of the vacuum manifold 
($\Pi_2({\rm SU}(N)/{\rm U}(1)^{N-1}) = Z^N$). The
magnetic field of such a monopole is long range. \be B_\mu =
{x^\mu \over 4 \pi r^3} \vec{g} \cdot \vec{H}. \ee The $N-1$
dimensional vectors $\vec{g}$ are determined by the 
non-Abelian
generalization of the Dirac quantization 
condition\,\cite{englert,goddard}
\be e^{i g\vec{g}\cdot \vec{H} }= I.
 \ee Solutions of this
quantization condition take the form \be \vec{g}= {4\pi 
\over g}
\sum_{i= 1}^{N-1} n_i \vec{\beta}^*_i \ee where 
$\vec{\beta}^*$
are the dual roots defined by $\vec{\beta}^* = \vec{\beta}/
|\vec{\beta}|^2$. We will be working with roots normalized 
to
unity, and thus $\vec{\beta}^* = \vec{\beta}$. The integers 
$n_i$
are elements of the group $\Pi_2$ (see Ref. 
\citebk{weinberg}). The 
monopoles
which have the smallest action correspond to roots taken 
once. The
action of these
 monopoles in the BPS limit is
\be M_{\vec\alpha} = {4 \pi \over g}\vec{h} \cdot 
\vec{\alpha}=
{4\pi M_{W_{\vec\alpha}}\over g^2}. \ee Just like with $W$ 
bosons
we can think of monopoles corresponding to simple roots as
fundamental ones with magnetic charges and action \be
 \vec{g_i}= {4\pi \over g}\vec{\beta}_i  \hskip 1 cm
 M_i = {4 \pi \over g}\vec{h} \cdot \vec{\beta}_i.
\ee
 For example, in the case of SU$(3)$ (see 
Eqs.~(\ref{simpleroots},\ref{wmass}))
the monopole action  spectrum (in the BPS limit) is
\be
M_1= {2\pi \over g}(h_1 +\sqrt{3} h_2 ), \hskip 0.5 cm M_2= 
{2\pi\over g}(h_1 -
\sqrt{3} h_2 ), \hskip 0.5 cm
M_3= {4\pi\over g}h_1.
\ee

The effect of these monopoles is to impart finite mass to 
all the
perturbatively massless ``photons.'' The derivation of the
effective Lagrangian follows exactly the same lines as the
original derivation of Polyakov\,\cite{polyakov} for the 
SU$(2)$ 
theory.
The resulting low-energy effective theory is
written in terms of the $N-1$ component field, $\vec{\eta}$, 
with
the following Lagrangian:\,\cite{wadia,snyderman} \be
{\cal{L}}_{\rm eff} = {g^2\over 32\pi^2} 
(\d_{\mu}\vec{\eta})^2 
+
\sum_\alpha {M_\alpha^2 g^2\over 16\pi^2 
}\mbox{exp}(i\vec{\alpha}
\cdot \vec{\eta}) .\label{lowenergy} \ee The sum is over all
$N(N-1)$ non-vanishing roots. The potential induced by the
monopoles is proportional to the monopole fugacity 
\be 
M^2_\alpha=
{16 \pi^2 \xi_{\alpha}\over{g^2}}\,, \qquad \xi_{\alpha} 
=
\mbox{const.}\,\, {M_{W_\alpha}^{7/2}\over g} \exp\left[{-
{4\pi 
M_{W_\alpha}
\over g^2} \epsilon\left({M_H\over M_W}\right)}\right]\,. 
\ee
At weak coupling the photons are much lighter than the $W$  
bosons and thus are the only relevant degrees of freedom in 
the
low-energy sector.

\subsection{The magnetic $Z_N$ symmetry}

The global symmetry structure is very important for the 
understanding of the
deconfining transition. The relevant symmetry in the present 
model 
is the
magnetic $Z_N$ symmetry. We now wish to explain how this 
symmetry is
implemented in the effective low-energy Lagrangian  ---  
generalizing 
the result
 obtained for $Z_2$ in the previous section.

The order parameter of the magnetic symmetry is the set of
magnetic vortex operators $V_i$, $i=1,...,N-1$. These 
operators
were constructed explicitly in Ref. \citebk{Kovner}. These 
operators
carry the magnetic fluxes of the $N-1$ U$(1)$ Abelian 
magnetic
fields. The defining commutation relation for $V_i$ is \be
[V_i(x),{\vec B}(y)]=-{4\pi\over g}{\vec 
w}_iV_i(x)\delta^2(x-y).
\label{nvortex} \ee 
Here ${\vec B}$ is the $N-1$ dimensional
vector of magnetic fields,\footnote{These magnetic fields 
can be
constructed in an explicitly gauge invariant way from the
non-Abelian field strengths and the Higgs field, see Ref.
\citebk{Kovner}.} whose $j$-th component is the projection 
of the
non-Abelian field strength onto the direction of the Cartan
subalgebra generator $H_j$, and ${\vec w}_j$ are $N-1$ 
weight
vectors of SU$(N)$. The choice of the $N-1$ out of $N$ 
weight
vectors is arbitrary. A change in this choice will lead to 
the
redefinition of the vortex operators such that the new 
operators
will be products of the old ones and their conjugates. It is
always possible to choose these weights so that together 
with the
``fundamental'' roots $\beta_i$ they satisfy the relation 
\be
{\vec w_i}{\vec \beta_j}={1\over 2}\delta_{ij}. 
\label{relat} \ee
The flux eigenvalues in Eq.~(\ref{nvortex}) are dictated by 
the
requirement of the locality of the vortex operators and are
analogous to the Dirac quantization condition. The explicit 
form
of the vortex operators in terms of the field $\eta$ in
Eq.~(\ref{lowenergy}) is \be V_i(\vec{x})= {g\over\sqrt{8}
\pi}\,e^{i\chi_i} \ee with \be \chi_i =  \vec{w}_i \cdot
\vec{\eta}  \Longrightarrow \vec{\eta} = 2 \sum_i 
\vec{\beta}_i
\chi_i .\label{chieta} \ee The effective Lagrangian can be 
written
as a nonlinear $\sigma$-model in terms of $V_i$ as \bea
{\cal{L}}_{\rm eff} = {N-1\over 2}\sum_{i,j} A_{ij} {1\over
V_k^*V_k}(V_i^*\d_{\mu}V_i)(V_j \d_{\mu}V_j^*) + \nonumber
\\
\lambda(\sum_i
(V_i V_i^* - {g^2\over 8\pi^2})^2 + \sum_{\alpha}
k_{\alpha}\,\prod_i V_i^{2 i\vec{\alpha} \cdot 
\vec{\beta}_i}
\label{vlag}
\eea
 with $\lambda\rightarrow\infty$.

The matrix $A_{ij}= 2\vec{\beta}^*_i \cdot \vec{\beta}_j$ 
depends
on the choice of the fundamental roots. With the 
conventional
choice of positive roots, where $\vec\beta_i\vec\beta_j=-
1/2, \ \
\ i\ne j$, it is the Cartan matrix of the Lie algebra. All 
its
diagonal elements are equal to $2$, while all its off 
diagonal
elements equal $-1$. We will find it however more convenient 
in
the following to use a different set of fundamental roots, 
for
which $\vec\beta_i\vec\beta_j=1/2, \ \ \ i\ne j$. Such a 
choice is
always possible for any SU$(N)$. For this choice of roots 
the
off-diagonal matrix elements of $A_{ij}$ are all equal to 
$1$.

For SU$(3)$ we have
\be
{\bf A}=\left ( \begin{array}{cc}
2 & 1\\
1& 2
\end{array}
\right )
\ee
and the effective Lagrangian
\bea
{\cal{L}}_{\rm eff}=  &&\d_{\mu}V_1 \d_{\mu}V_1^* +
{8\pi^2\over g^2}V_1^*\d_{\mu}V_1 V_2 \d_{\mu}V_2^* +
\d_{\mu}V_2 \d_{\mu}V_2^*
\nonumber \\[2mm]
&&+ \xi_1( V_1 V^*_2 + \mbox{c.c} ) + \xi_2( V_1^2 V_2 +
\mbox{c.c} ) +\xi_3( V_1 V_2^2 + \mbox{c.c} ). 
\label{vortexlag}
\eea

The magnetic $Z_N$ symmetry has an obvious and simple 
representation 
in this
effective Lagrangian as $$V_i\rightarrow\exp\{2\pi 
in/N\}V_I\,.$$

As long as only small fluctuations of the phase fields 
$\chi_i$
are important, the Lagrangian (\ref{vortexlag}) is 
equivalent
to Eq.~(\ref{lowenergy}). Thus, at low temperature the 
descriptions
based on these Lagrangians are equivalent. The difference 
appears
just as for SU$(2)$ only when the phase nature of $\chi_i$ 
plays a
role. Indeed, since $\chi_i$ are treated in 
Eq.~(\ref{vortexlag})
as phases, dynamically one allows configurations in which 
these
phases have nontrivial winding. On the other hand, in
Eq.~(\ref{lowenergy}) such configurations cost an infinite 
amount
of energy. As discussed in detail in Refs. \citebk{Kovner} 
and
\citebk{dunne} the winding configurations correspond to the 
heavy
$W$ bosons. In fact, the explicit relation between the 
vorticity of
the fields $V_i$ and the electric charges is given by
\cite{Kovner} \be {1\over g}{\vec w}_i{\vec Q}={1\over
4\pi}\oint_{C\rightarrow\infty}dl_\mu
\partial_\mu\chi^i_\mu .
\label{vorticity} \ee Thus, the difference between the two
Lagrangians is important whenever the physics of the $W$ 
bosons
plays an important role. We have seen in the case of the 
SU$(2)$
theory that $W$'s are indeed important near the phase 
transition
temperature. The same turns out to be true for arbitrary 
SU$(N)$.
We thus have to be careful to treat the $W$   bosons 
properly in
the transition region.\cite{suN} 

\subsection{The deconfining transition}
Close to the transition temperature we can safely use the
dimensionally reduced Lagrangian. The zero Matsubara 
frequency
sector is described by the {\it two}-dimensional Lagrangian 
\be
{\cal{L}}_{\rm eff} = {g^2\over 32\pi^2 T} 
(\d_{\mu}\vec{\eta})^2 +
\sum_\alpha {M_\alpha^2 g^2\over 16\pi^2 T
}\mbox{exp}(i\vec{\alpha} \cdot \vec{\eta}) .\label{reduced} 
\ee
However, as we noted before, our description should include 
$W$
bosons, and so the fields $\eta$ should be treated here as 
phases
with periodicity appropriate to Eq.~(\ref{chieta}). In fact, 
the
Lagrangian also has to be augmented by a four derivative
``Skyrme'' term, which fixes the energy of the winding 
states to
be equal to the masses of $W$ bosons.\cite{dunne}  Things 
are
however simplified once we note that the density of $W$ 
bosons at
criticality is exponentially small due to the Boltzmann 
factor
suppression. Thus, $W$'s can be treated in the dilute gas
approximation in the same way as in the SU$(2)$ theory. To 
do this
explicitly we first have to understand how to write the 
partition
function in the presence of one $W$ boson of a particular
type.\cite{suN} 

Let us first consider a $W$ boson corresponding to one of 
the
fundamental roots $\beta_k$. Combining Eq.~(\ref{cham}) with
 (\ref{relat}) and (\ref{vorticity}) we see that this $W$
boson corresponds to unit vorticity of the field $V_k$ and 
zero
vorticity of all other fields $V_j$, $j\ne k$. To create 
such a
vortex in the path integral we must introduce an external
``current''which forces the discontinuity of the field 
$\chi_k$
\be \chi_k = \chi_k + 2\pi .\ee The partition function in 
the
presence of one $W$ boson is thus \bea
 Z = \int D[\chi(x)]
\exp\left\{-\int d^2 y
\left( S[A] +\sum_{\alpha}\zeta_{\alpha} \cos (2\sum_i 
\vec{\alpha}\cdot
\vec{\beta}_i \chi_i) \right) \right\}\,,
\eea
where
\bea
S[A] = {g^2 \over 16 \pi^2 T}\sum A_{ij}
(\d_{\mu}\chi_i- J^i_\mu(y,x))(\d_{\mu}\chi_j- 
J^j_\mu(y,x))\,,
  \label{onev1}
\eea
where
\be J^i_{\mu}(y,x)=2\pi\delta_{ik} n_\mu(y)\delta(y\in C_x) 
\ee
with $C_x$ a curve that starts at the location of the vortex 
(the
point $x$), and goes to infinity, and $n_\mu$ is the unit 
normal
to this curve. The insertion of this current forces the 
normal
derivative of $\chi_k$ to diverge on curve $C$, so that 
$\chi_k$
jumps by $2\pi$ across $C$. Since in the rest of the space
$\chi_k$ is smooth, the path integral is dominated by a
configuration with unit vorticity of $\chi_k$.

The path integral Eq.~(\ref{onev1}) differs from the 
partition
function in the vacuum sector by the linear term in the 
Lagrangian
\be - { g^2 \over 4 \pi^2 T } \sum_{i,j}\int d^2 y\,
\vec{\beta}_i\cdot \vec{\beta}_j\, \partial_{\mu}\chi_i 
J^j_{\mu}
= - { g^2 \over 4 \pi T } \int_{C_x}
dx_\mu\epsilon_{\mu\nu}\,\partial_\nu\vec{\beta_k} \cdot 
\vec\eta
. \ee Defining in the standard way the dual field 
$\tilde\eta$,
\be i\d_{\mu}\vec{\tilde{\eta}} = \epsilon_{\mu \nu}
\d_{\nu}\vec{\eta}\,,
 \ee 
we can recast the contribution of 
this
particular $W$ boson in the form of the following extra term 
in the Lagrangian
 \be - i{ g^2 \over 4 \pi T 
}\vec{\beta_k}\cdot
\vec{\tilde{\eta}} \,.\label{wcont} 
\ee
 This procedure can be
repeated for a $W$ boson corresponding to an arbitrary root
$\alpha$ with the only difference that in Eq.~(\ref{wcont}), 
the
root $\beta_k$ is replaced by the root $\alpha$. To create 
several
$W$ bosons one just inserts the external current which is 
the sum
of the currents creating individual $W$'s.

A dilute ensemble of such objects with small fugacities
$\mu_{\alpha}$ then has \be Z= \prod_{\alpha} 
\sum_{n,m}{1\over
n!}{1\over m!}\mu_{\alpha}^{n+m}\int \prod_idx_i\prod_jdy_j
Z(x_i,y_j). \ee  The result of the summation over the number 
of
$W$'s is the partition function with the Lagrangian 
\be
{\cal{L}}_{\rm eff} = {g^2\over 32\pi^2 T} 
(\d_{\mu}\vec{\eta})^2 +
\sum_\alpha \zeta_{\alpha}\mbox{exp}(i \vec{\alpha} \cdot
\vec{\eta}) +\sum_\alpha \mu_{\alpha}\mbox{exp}\left(i {g^2\over 
4\pi
T} \vec{\alpha} \cdot \vec{\tilde{\eta}}\right) 
\label{lowenergy2} 
\ee
with summation in both terms going over all non-vanishing 
roots of
SU$(N)$. The coefficients $\mu_\alpha$ are proportional to 
the
fugacities of the corresponding $W$ bosons \be 
\mu_\alpha\propto
\exp\{-M_{W_\alpha}/T\} . \ee
 Equation (\ref{lowenergy2}) is the
dimensionally reduced theory which we will now use to study 
the
phase transition.

Let us first disregard the $W$ boson induced term in the 
effective
Lagrangian. If we do that,
 we are back to
the theory of Eq.~({\ref{reduced}). An interesting property 
of this
theory is that since the group is simply laced (all the 
roots are
of unit length) the anomalous dimensions of all the 
interaction
terms are equal. The scaling dimension of all the monopole 
induced
terms is \be \Delta_M = {4\pi T\over g^2} .\ee This 
immediately
tells us that at the temperature \be T_{\rm BKT}= {g^2\over 
2\pi} ,\ee
all these interactions become irrelevant. Thus, at $T_{\rm 
BKT}$ 
one
 expects the Berezinsky-Kosterlitz-Thouless transition to 
take 
place.
Above this temperature the infrared behavior of the theory 
is
that of $N-1$ free massless particles. Note that $T_{\rm 
BKT}$ 
does
not depend on the number of colors $N$. If the picture just
described were true, the universality class of the phase
transition would be that of U$^{N-1}(1)$.

Of course, this is exactly the same situation as encountered 
in Ref.
\citebk{zarembo} in the SU$(2)$ case. Again just like in 
SU$(2)$
case this conclusion is incorrect due to the contribution of 
the
$W$ bosons. To see this, it is simplest to ask what would 
happen
at high temperature if there were no monopole contributions 
at
all. This amounts to studying Eq.~(\ref{lowenergy2}) with
$\xi_\alpha=0$. This theory describes non-compact 
electrodynamics
with $N-1$ photons and the spectrum of charged particles 
given by
Eq.~(\ref{cham}). This limit is again simple to understand, 
since
the theory is exactly dual to the theory with monopoles and
without charges. The scaling dimensions of all the $W$ 
induced
perturbations are equal and are given by \be \Delta_W = 
{g^2\over
4\pi T} .\ee Thus, the perturbations are irrelevant at low
temperature, but become relevant at \be T_{NC}={g^2\over 
8\pi}
.\ee Since $T_{NC}<T_{\rm BKT}$ this tells us that we can 
not 
neglect
the effects of charges at criticality. The story of SU$(2)$
exactly repeats itself. Even the value of the temperature at 
which
the scaling dimensions of the charge ---  and monopole 
induced
perturbations are equal does not depend on $N$.

We expect therefore that the actual transition temperature 
is \be
T_C={g^2\over 4\pi}\,, \ee at which point all perturbations 
have the
same scaling dimension. This expectation is supported by the
renormalization group analysis.

The renormalization group equations for the theory
Eq.~(\ref{lowenergy2}) were studied in Ref. 
\citebk{boyanovsky}. In
general the equations are quite complicated due to the cross
correlations between different operators. For this reason
 the space of
parameters of the theory has to be enlarged if one wants to 
study
the flow whose ultraviolet initial condition is provided by
Eq.~(\ref{lowenergy2}) with arbitrary values of fugacities. 
However,
there is one simple case, that is when the initial condition 
is
such that all the monopole fugacities are equal
$\xi_{\alpha_i}=\xi_{\alpha_j}=\xi$, and all the charge 
fugacities
are equal $\mu_{\alpha_i}=\mu_{\alpha_j}=\mu$. This initial
condition is stable under the renormalization group flow. In 
this subspace the 
renormalization group
equations, written in terms of the scaled temperature 
$t={4\pi
T\over g^2}$ and dimensionless fugacities, read \bea
&&{\d t\over \d \lambda} = 2 \pi^2 N t( \mu^2 - \zeta^2 ) 
,\\
&& {\d \mu \over \d \lambda} = (2 - {1\over t}) \mu  - 2 
\pi(N-2) 
\mu^2 ,\\
&&{\d \zeta \over \d \lambda} = (2 -  t) \zeta  - 2 \pi(N-2)
\zeta^2 .\label{rge} \eea These equations have exactly the
property reflecting our previous discussion. That is the 
points
$t=2, \ \ \mu=0$ and $t=1/2, \ \ \ \xi=0$ are both unstable. 
The
stable infrared fixed point is \be t_0 = 1\hskip 1 cm \mu_0= 
\zeta_0 =
{1 \over 2\pi(N-2)} \,.\label{fixedp} \ee One can in fact 
easily
check that, in the three-dimensional space of couplings 
$t,\xi$
and $\mu$, this point has two attractive and one repulsive
direction. This is precisely what one expects from the 
infrared 
fixed
point located on the critical surface, the two attractive
directions being the tangential directions to the surface.

The renormalization group equations have an obvious duality 
symmetry,
$\mu\rightarrow\xi, \ \ \ t\rightarrow 1/t$. This is the
reflection of the transformation $\eta\rightarrow\tilde\eta$ 
on
the level of the Lagrangian (\ref{lowenergy2}). The 
points
$t=1$, $\mu=\xi$ are symmetric under duality, and this 
ensures the
existence of a self dual fixed point. This is important, 
since the
exact position of the fixed point is scheme dependent. Its
existence however is ensured by the duality symmetry.

What is the nature of this fixed point? For $N=2$ we were 
able in
Ref. \citebk{dunne} to fermionize the fixed point theory and 
show
explicitly that it is equivalent to one massless Majorana 
fermion.
We are not able to perform a similar analysis for arbitrary 
$N$.
There are however several comments that we would like to 
make.
Phase transitions in $Z_N$ invariant spin models have been 
studied
quite extensively. A recent discussion of the situation is 
given
in Ref. \citebk{znspin}. One considers a spin model of one 
phase 
field
$\theta$ with a symmetry breaking term of the type $h\
\cos\{N\theta\}$ which breaks the U$(1)$ symmetry down to 
$Z_N$.
When the coefficient $h$ of this symmetry breaking term is 
large,
the model resembles the Potts model and thus (for $N>4$) has 
a
first order phase transition. When the breaking is small on 
the
other hand, the behavior is similar to the Villain model: 
the
system undergoes two BKT type transitions with a massless 
U$(1)$
symmetric phase at intermediate temperatures. At some 
particular
``tricritical'' value of $h$ the massless phase shrinks to a 
point
and it comes together with the first order transition line. 
This
tricritical point is self-dual and is described by a 
conformal
$Z_N$ invariant parafermionic theory with the central charge
$c=2(N-1)/(N+2)$ introduced in Ref. \citebk{zamolodchikov}. 
In this 
type
of model therefore one generically expects either a first 
order
transition or a pair of BKT transitions with a massless 
phase in
between. The tricritical behavior is special and requires 
fine
tuning of the parameters. This is indeed also the prevailing
general expectation for the order of the transition in 2+1
dimensional gauge theories at large $N$: either first order 
or
Villain type U$(1)$ invariant behavior.

In fact, we find in our model a completely different 
situation. The
transition is not first order, and there is no U$(1)$ 
invariant
massless phase. We stress that within the renormalization 
group flow 
Eq.~(\ref{rge})
the infrared fixed point Eq.~(\ref{fixedp}) has two 
attractive
directions. This means that it governs the infrared behavior 
of 
the
points which lie on a two-dimensional critical surface in 
the
three-dimensional parameter space, and is therefore generic. 
This
by itself does not preclude the possibility that this fixed 
point
is the same as in the parafermionic $Z_N$ theory of Ref. 
\citebk{zamolodchikov}. If this is the case, it is quite
interesting, since the point which appeared as 
``tricritical''
from the point of view of usual spin models is in fact 
generic
from the point of view of the 3D gauge theories. Although we 
can
not prove
 that our critical point is
described by the parafermionic theory, we will present some
arguments
 supporting this conjecture.
 The point is that, as
opposed to models considered in Ref. \citebk{znspin} our 
Lagrangian (\ref{lowenergy2}) describes a theory of $N-1$ 
light  fields.
The theory of $N-1$ free massless fields has the ultraviolet 
central 
charge
$c_{\rm UV} = N-1$. However, this conformal field theory is 
deformed by the 
monopole and
$W$-induced perturbations and flows to a different infrared 
fixed
point. However, let us note that the central charge $c=N-1$ 
is
precisely the central charge of the SU$(N)_1$ 
Wess-Zumino-Novikov-Witten (WZNW) model. 
The
Ising (i.e. $c=1/2$) model  is the lowest among the minimal 
models
with Virasoro (i.e. $W_2$)  symmetry. The highest model of 
this
class is the $c=1$ model (one free field ) which is 
precisely the
 SU$(2)_1$ WZNW model.
When the $c=1$ model was deformed by the monopole and 
$W$-boson
operators the central charge was reduced  ---  and the 
resulting infrared
theory was  Ising-like.

Now,  $Z_N$ parafermions with $c = 2(N-1)/N+2$  are the 
lowest
minimal
 models with $W_N$ symmetry, and the highest is
SU$(N)_1$  (for more information about parafermions see, for
example, Ref.   \citebk{cappelli} and references therein), 
which can 
be
described  in terms  of
 $N-1$  massless  fields.
Thus, if the theory in the ultraviolet describes $N-1$  
massless 
fields  and
has $W_N$ symmetry, it is quite possible  that result of the
relevant (monopole+$W$)  deformation is a self-dual critical
point. It is indeed known that the $Z_N$ parafermion theory 
is the
self-dual model with $W_N$ symmetry. The fact that the 
central
charge (and thus the effective number of degrees of freedom) 
is
reduced in the process of the flow towards the infrared is 
of 
course in
complete accord with Zamolodchikov's $C$-theorem. It is 
therefore
possible that the infrared fixed point that describes the 
universality
class of the Georgi-Glashow model is the conformal $Z_N$ 
parafermion 
theory.

Analysis of Ref. \citebk{boyanovsky}, although admittedly 
incomplete 
also
supports the expectation that we do not have the Villain 
picture.
As discussed in detail in Ref. \citebk{suN}, it is the 
presence of 
the
large number of fields that drives our theory away from the
Villain behavior.

 An interesting feature of this result is that the
critical temperature in the SU$(N)$ theory at large $N$ is
proportional to the coupling $g^2$ and not to the 't Hooft
coupling $\lambda=g^2N$. Thus, at large $N$ the critical
temperature approaches zero. The physical reason for this is 
easy
to understand. At large $N$ and fixed $\lambda$ the Higgs 
VEV
should also scale with $N$ in such a way that the mass of 
$W$
bosons remains fixed. The monopole action then grows as $N$ 
and
the photons get progressively lighter (exponentially with
$N$)\footnote{This is analogous to the situation in QCD 
where the
instantons become less relevant at large $N$ and the 
$\eta^\prime$
meson becomes massless. The major difference is of course 
that
while the $\eta^\prime$ mass in QCD decreases as $1/N$, the 
photon
masses in Georgi-Glashow model decrease exponentially. This 
difference 
is due
to the non diluteness of the instanton gas in QCD as opposed 
to
diluteness of the monopole gas in the Georgi-Glashow 
model.}. Thus, the
thickness of the confining string grows and the density of 
$W$
bosons needed to restore the symmetry becomes smaller and 
smaller.

More importantly, our main conclusion is that the 
deconfining
transition in the SU$(N)$ Georgi-Glashow model is second 
order and the
universality class is determined by the infrared fixed point
Eq.~(\ref{fixedp}). This point is $Z_N$ symmetric and self 
dual. We
have given some arguments supporting the possibility that  
the
fixed point theory is the $Z_N$ parafermionic 
model\,\cite{zamolodchikov} although  we are not able to 
prove this
explicitly. We can however definitely exclude the Potts and
Villain universality classes.

\section{Concluding remarks}
We have discussed at length various aspects of the confining
physics and the deconfining phase transition in weakly 
interacting
gauge theories in 2+1 dimensions. Naturally one should ask: 
where
do we go from here? The ultimate goal we have in mind is of 
course
strongly interacting 3+1 dimensional gauge theories. But 
this goal
is still very far away. Many aspects of our discussion would
change with dimensionality. In 3+1 dimensions monopoles are 
not
instantons any longer, but particles, while the magnetic 
vortices
are not particles but strings. The analog of the effective 
low-energy theories we have discussed, which incorporate the 
effects
of these objects is not known. Moreover, weakly coupled 
phases of
gauge theories are generically not confining and separated 
from
the confining ones by phase transition. Thus,  even if such
effective theories were known it is not clear to what extent 
they
can be used as a guide for the physics of the confining 
phase. The
situation is not hopeless though, and a lot of work is being 
done
on 3+1 dimensional theories. In particular, the dual
superconductivity picture within the Abelian dominance 
hypothesis
has been studied quite extensively.\cite{abelian} More 
recently
the vortex-based ideas have also gained support.\cite{zn}

Perhaps a simpler question is what can one say about the 
strongly
interacting regime in 2+1 dimensions. Here some progress has 
been
made. Both monopole based\,\cite{das} and vortex based 
approaches\,\cite{Kovner2} have been advanced. The basic 
encouraging 
feature
is that the lightest degrees of freedom seem to be the same 
in the
strongly and weakly coupled regimes. This also seems to be 
the
case near the deconfining transition temperature.\cite{suN}  
Thus,
a certain continuity is there and one hopes eventually to 
learn
how to utilize it to the full extent.

In short, the problems are hard  ---  but don't despair ...

\section*{ Acknowledgments} We thank our collaborators 
Cesar
Fosco, Chris Korthals Altes, Baruch Rosenstein,  Martin
Schvellinger, Ben Svetitsky and Bayram Tekin for 
collaboration and
discussions on subjects covered in this review. We thank Ben
Gripaios for reading the manuscript. This work has been 
supported 
partly
by PPARC. \vskip 1cm


{\arabic{enumi}.}
\renewcommand{\baselinestretch}{1.2}



\begin{thebibliography}{99}


\bibitem{polyakov} A.M. Polyakov, 
\Journal{\PLB}{59}{82}{1975};
 \Journal{\NPB}{120}{429}{1977}.


\bibitem{lattice} G. Bhanot and B.A. Freedman,
\Journal{\NPB}{190}{357}{1981};
Y. Munehisa, \Journal{\PLB}{155}{159}{1985};
G.N. Obodi, \Journal{\PLB}{174} {208} {1986};
M.G. Amaral and M.E. Pol, {\it J. Phys. G:
Nucl. Part. Phys.} {\bf 16}, 1 (1990).

\bibitem{kovnerr} A. Kovner, in M. Shifman, (Ed.)
{\em At the 
Frontier of Particle 
Physics}, Handbook of QCD/Ioffefestschrift,
(World Scientific, Singapore, 2001)
Vol. 3, pp. 1777 - 1825 [hep-ph/0009138].

\bibitem{frsh} E. Fradkin and S.H. Shenker,
\Journal{\PRD}{19}{3682}{1979}.

\bibitem{Prasad}
M. K. Prasad and C. M. 
Sommerfield,\Journal{\PRL}{35}{760}{1975};
 E. B. Bogomolny, Sov. J. Nucl. Phys. {\bf 24}, 861 (1976);
T. W. Kirkman and C. K. Zachos, 
\Journal{\PRD}{24}{999}{1981}.


\bibitem{Coleman}
S. Coleman, \Journal{\PRD}{11}{2088} {1975};
S. Samuel,  \Journal{\PRD}{18}{1916} {1978};
 D. J. Amit, Y. Y. Goldschmidt and G.
Grinstein, {\em J. Phys. A 15} 585 (1980);
D. Boyanovsky, {\em J. Phys. A 22} 2601 (1989).

\bibitem{book}   A.M. Polyakov, {\it Gauge Fields and 
Strings},
 (Harwood Academic Publishers, Chur, Switzerland, 1987).

\bibitem{'t Hooft}
G. 't Hooft, \Journal{\NPB}{138}{1}{1978}, Acta Phys. Austr.
Suppl. XXII (1980) 531.

\bibitem{samuel}S. Samuel, \Journal{\NPB}{154}{62}{1979}


\bibitem{Kovner}A. Kovner and B. Rosenstein,
\Journal{\IJMP}{7}{7419}{1992}.


\bibitem{Kovner1}C.P. Korthals-Altes and A. Kovner,
\Journal{\PRD}{62}{ 096008} {2000}


\bibitem{Kovner2} A. Kovner and B. Rosenstein,
 \Journal{\JHEP}{003}{9809}{1998}.


\bibitem{greensite} J. Ambjorn and J. Greensite,
 \Journal{\JHEP}{004}{9805}{1998}.

\bibitem{POL96}
A.M. Polyakov, \Journal{\NPB}{486}{23}{1997}, hep-
th/9607049.

\bibitem{confiningstringaction}
M.C. Diamantini, F.Quevedo and C.A. Trugenberger,
\Journal{\PLB}{396} {115} {1997} [hep-th/9612103];
D. Antonov, \Journal{\PLB}{427} {274} {1998} 
[hep-th/9804016];
 \Journal{\PLB}{428} {346} {1998} [hep-th/9802056].


\bibitem{Rigid1}
A.M. Polyakov, \Journal{\NPB}{268}{406}{1986};
H. Kleinert, \Journal{\PLB}{174} {335} {1986}.


\bibitem{confinigstringreferences}
M.C. Diamantini, H.Kleinert,  C.A. Trugenberger,
\Journal{\PLB}{457} {87} {1999}, hep-th/9903208

\bibitem{KALB}
V.I. Ogievetsky, I.V. Polubarinov, {\em Sov. J. Nucl. Phys.} 
{\bf 
4}
156 (1967), {\em Yad. Fiz.} {\bf 4} 216  (1966);
 M. Kalb, P. Ramond, \Journal{\PRD}{9}{2273} {1974};
D.Z. Freedman, P.K. Townsend, 
\Journal{\NPB}{177}{282}{1981}.


\bibitem{martin} I.I. Kogan, A. Kovner and M. Schvellinger
 \Journal{\JHEP}{0107}{019}{2001} [hep-th/0103235].

\bibitem{fosco}C. Fosco and A. Kovner,
 \Journal{\PRD}{63}{045009} {2001} [hep-th/0010064].

\bibitem{var} I. Kogan and A. Kovner
 \Journal{\PRD}{51}{1948} {1995} [hep-th/9408081].

\bibitem{ben} A. Kovner and B. Svetitsky,
 \Journal{\PRD}{60}{105032} {1999} [hep-lat/9811015].

\bibitem{vort}
E.C. Marino,  \Journal{\PRD}{38}{3194} {1998};
A. Kovner, B. Rosenstein and D. Eliezer,
\Journal{\NPB}{350}{325}{1991}.

\bibitem{DQSW} S. D. Drell, H. R. Quinn, B. Svetitsky, and 
M. 
Weinstein,
 \Journal{\PRD}{19}{619} {1979}.


\bibitem{zarembo}
N.~O.~Agasian and K.~Zarembo,
 \Journal{\PRD}{57}{2475} {1998} [hep-th/9708030].
For the SU$(N)$ theory similar analysis
omitting the effects of $W$ bosons was performed recently in
N.O. Agasyan and D. Antonov, hep-th/0104029.

\bibitem{dunne}
G.~Dunne, I.~I.~Kogan, A.~Kovner and B.~Tekin,
 \Journal{\JHEP}{0101}{032}{2001} [hep-th/0010201].

\bibitem{chernodub} M.N. Chernodub, E.-M. Ilgenfritz, A. 
Schiller
 \Journal{\PRD}{64}{054507}{2001};
 \Journal{\PRD}{64}{114502}{2001} [hep-lat/0106021].

\bibitem{BKT}
V.L. Berezinskii, Sov. Phys. JETP {\bf 32} 493 (1971);
J.M. Kosterlitz and D.J. Thouless, J. Phys. {\bf C7}  1181 
(1973).

\bibitem{yaffe} B. Svetitsky and L.G. Yaffe,
\Journal{\NPB}{210}{423}{1982}.

\bibitem{Kadanoff}
J.V. Jos\'{e}, L.P. Kadanoff, S. Kirkpatrick and D.R. Nelson,
\Journal{\PRD}{16}{1217}{1977};\\
A. P. Young,  {\em J. Phys.}  {\bf C11}  L4553 (1978);\\
K. Huang and J. Polonyi, \Journal{\IJMP}{6}{409}{1991}.

\bibitem{Nersesyan}
D.G. Shelton, A.A. Nersesyan and A.M. Tsvelik,
 \Journal{\PRB}{64}{8521}{1996}.


\bibitem{Tsvelik} A.O.Gogolin, A.A. Nersesyan and A.M. 
Tsvelik, {\em 
Bosonization and Strongly Correlated Systems}, (Cambridge 
University
Press, 1998).

\bibitem{kaks} C.P. Korthals-Altes, A. Kovner and M. 
Stephanov,
 \Journal{\PLB}{469}{205}{1999}.


\bibitem{inst}I. I. Kogan, A. Kovner and B. Tekin,
 \Journal{\JHEP}{0103}{021}{2001} [hep-th/0101171].


\bibitem{HAGEDORN} R. 
Hagedorn,~\Journal{\NCA}{3}{147}{1965}.

\bibitem{FVHW}
S. Fubini and G. Veneziano,~\Journal{\NCA} 
{64A}{1640}{1969};\\
K. Huang and S. Weinberg,~\Journal{\PRL}{25}{895}{1970}.

\bibitem{FC}
S. Frautschi,~\Journal{\PRD}{3}{2821}{1971}; \\
R.D. Carlitz,~\Journal{\PRD}{D5}{3231}{1972}.

\bibitem{cab}
N. Cabibbo and G. Parisi,~\Journal{\PLB}{59}{67}{1975}.


\bibitem{BOOKS} J. Polchinski,
{\em String Theory}, Vols. 1, 2, (Cambridge 
University
Press, 1998).

\bibitem{ABKR}
S.A. Abel, J.L.F. Barb\'on, I.I. Kogan and E. Rabinovici,
 \Journal{\JHEP}{9904}{015}{1999} [hep-th/9911004].

\bibitem{KOGAN87}
I.I. Kogan, {\em JETP Lett.} {\bf 45} 709 (1987) [{\em Pisma 
Zh. Eksp. 
Teor.
Fiz.} {\bf 45} 556  (1987)].

\bibitem{SATHIAPALAN87}
B. Sathiapalan,
 \Journal{\PRD}{35}{3277}{1987}.


\bibitem{ATICK}
J.J. Atick, E. Witten,
\Journal{\NPB}{310}{291}{1988}.

\bibitem{AK}
A.A. Abrikosov, Jr., I.I. Kogan,
 Sov.Phys.JETP {\bf 69} 235 (1989);
 \Journal{\IJMP}{1501}{1991}.

\bibitem{SATH}
B. Sathiapalan,
\Journal{\MPLA}{13}{2085}{1998} [hep-th/9805126].


\bibitem{hotsoup}
 D.A. Lowe, L. Thorlacius,
 \Journal{\PRD}{51}{665}{1995}
 [hep-th/9408134];
S. Abel, K. Freese, I. I.Kogan,
 \Journal{\JHEP}{0101}{039}{2001}
[hep-th/0005028].

\bibitem{KW}
I.I. Kogan, J.F Wheater,
\Journal{\PLB}{403}{31}{1997} [hep-th/9703141].

\bibitem{englert}
F.~Englert and P.~Windey,
 \Journal{\PRD}{14}{2728}{1976}.

\bibitem{goddard}
P.~Goddard, J.~Nuyts and D.~Olive,
\Journal{\NPB}{125}{1}{1977}.


\bibitem{weinberg}
E.~J.~Weinberg,
\Journal{\NPB}{167}{500}{1980}.

\bibitem{wadia}
S.~R.~Wadia and S.~R.~Das,
\Journal{\PLB}{106}{386}{1981}.


\bibitem{snyderman}
N.~J.~Snyderman, \Journal{\NPB}{218}{381}{1983}.


\bibitem{suN} I.I. Kogan, A. Kovner and B. Tekin,
\Journal{\JHEP}{0105}{062}{2001} [hep-th/0104047].

\bibitem{boyanovsky} D. Boyanovsky and R. Holman,
\Journal{\NPB}{358}{619}{1991}.


\bibitem{znspin} P. Dorey, R. Tateo and K.E. Thompson,
\Journal{\NPB}{470}{317}{1996} [hep-th/9501098.

\bibitem{zamolodchikov} A.B. Zamolodchikov and V.A. Fateev,
 Sov.Phys. {\em JETP} {\bf 62} 215 (1985);
 V.A. Fateev and A.B. Zamolodchikov,
\Journal{\NPB}{280}{644}{1987};
D.Gepner and Z. Qiu,
\Journal{\NPB}{285}{423}{1987}.


\bibitem{cappelli} A. Cappelli, L.S. Georgiev and 
I.T.Todorov,
\Journal{\NPB}{559}{499}{2001} [hep-th/0009229].

\bibitem{abelian} For a recent review see A. Di Giacomo,
hep-lat/0112002.

\bibitem{zn} M. Faber, J. Greensite and  S. Olejnik
\Journal{\JHEP}{0006}{041}{2000} [hep-lat/0005017].

\bibitem{das}  S. R. Das and  S. R. Wadia,
 \Journal{\PRD}{53}{5856}{1996} [hep-th/9503184].

\end{thebibliography}
\end{document}